\documentclass[a4paper,twocolumn,11pt,accepted=2024-06-28]{quantumarticle}
\pdfoutput=1

\usepackage[utf8]{inputenc} 
\usepackage[T1]{fontenc}    
\usepackage{hyperref}       
\usepackage{url}            
\usepackage{xurl}            
\usepackage{booktabs}       
\usepackage{nicefrac}       
\usepackage{microtype}      
\usepackage{graphicx}
\usepackage{xcolor, etoolbox}
\usepackage{amsmath,amsfonts,amsthm,amssymb} 
\usepackage{mathtools}
\usepackage{ragged2e}
\usepackage{braket}      
\usepackage[english]{babel} 
\usepackage{enumitem}
\usepackage{xspace}
\usepackage{dsfont}
\usepackage{bm}
\usepackage{bbm}
\usepackage{dcolumn}
\usepackage{mathrsfs}
\usepackage[sort&compress,numbers]{natbib}

\newcommand{\bTr}{\textbf{Tr}}
\newcommand{\Tr}{\text{Tr}}

\DeclarePairedDelimiter\abs{\lvert}{\rvert}

\begin{document}

\title{Entropic distinguishability of quantum fields in phase space}

\author{Sara Ditsch}
    \email{sara.ditsch@tum.de}
    \affiliation{Physik Department, TUM School of Natural Sciences, Technische Universität München, James-Franck-Straße 1, 85748 Garching, Germany}
    \affiliation{Max-Planck-Institut für Physik, Werner-Heisenberg-Institut, Boltzmannstr. 8, 85748 Garching, Germany}
    \orcid{0009-0002-0953-6656}
\author{Tobias Haas}
    \email{tobias.haas@ulb.be}
    \affiliation{Centre for Quantum Information and Communication, École polytechnique de Bruxelles, CP 165, Université libre de Bruxelles, 1050 Brussels, Belgium}
    \orcid{0000-0003-1477-9855}
    \thanks{\texttt{\href{https://tobi-haas.de}{tobi-haas.de}}}

\maketitle

\begin{abstract}
We present a general way of quantifying the entropic uncertainty of quantum field configurations in phase space in terms of entropic distinguishability with respect to the vacuum. Our approach is based on the functional Husimi $Q$-distribution and a suitably chosen relative entropy, which we show to be non-trivially bounded from above by the uncertainty principle. The resulting relative entropic uncertainty relation is as general as the concept of coherent states and thus holds for quantum fields of bosonic and fermionic type. Its simple form enables diverse applications, among which we present a complete characterization of the uncertainty surplus of arbitrary states in terms of the total particle number for a scalar field and the fermionic description of the Ising model. Moreover, we provide a quantitative interpretation of the role of the uncertainty principle for quantum phase transitions.
\end{abstract}\\

\vspace{-0.2cm}
TUM-HEP-1514/24
\vspace{-0.2cm}

\section{Overview}

\subsection{Introduction}
Uncertainty in incompatible measurements marks the dividing line between classical and quantum phenomena \cite{Heisenberg1927}. Besides the well-known formulations of the uncertainty principle in terms of second moments of measurement distributions \cite{Kennard1927,Weyl1928,Robertson1929,Robertson1930,Schroedinger1930}, entropic uncertainty relations (EURs) have gained interest over the past decades \cite{Everett1957,Hirschman1957,Beckner1975,Bialynicki-Birula1975,Deutsch1983,Kraus1987,Maassen1988,Berta2010,Frank2012} (see \cite{Wehner2010,Bialynicki-Birula2011,Coles2017,Hertz2019} for reviews). As EURs are typically tighter than second-moment relations, they proved to be crucial in many contexts, e.g., quantum key distribution \cite{Renes2009,Berta2010,Tomamichel2011,Furrer2012}, quantum cryptography \cite{Grosshans2004,Tomamichel2012}, entanglement witnessing \cite{Walborn2009,Walborn2011,Schneeloch2018,Schneeloch2019,Haas2021b,Haas2022a,Haas2022c,Haas2022d}, and quantum metrology \cite{Giovanetti2011,Hall2012}. 

Along with the rising importance of entropic descriptions for understanding field-theoretic phenomena \cite{Bombelli1986,Srednicki1993,Callan1994,Calabrese2004,Popescu2006,Calabrese2009,Casini2009,Islam2015,Kaufman2016}, the first formulation of an EUR for a quantum field has recently been put forward \cite{Haas2022b} (see also \cite{Haas2021a,Casini2020,Casini2021,Magan2021}). Therein, it was argued that any classical entropy $S$ over some measurement distribution diverges with the number of modes $N$, thereby rendering any EUR meaningless in the continuum. In contrast, a formulation in terms of a functional relative entropy, which serves as a measure for the \textit{entropic distinguishability} of distributions over field configurations, was shown to remain universally finite \cite{Haas2022b}. To this end, the uncertainty principle was expressed via a state-dependent upper bound on the entropic distinguishability of the distributions associated with the field $F[\phi] = \braket{\phi | \boldsymbol{\rho} | \phi}$ and the momentum field $G[\pi] = \braket{\pi | \boldsymbol{\rho} | \pi}$ with respect to their vacuum expressions $S [F \| \bar{F}] + S [G \| \bar{G}] \le B$, showing that distributions over incompatible fields can \textit{not} be distinguished arbitrarily well.

However, the relative entropic uncertainty relation (REUR) presented in \cite{Haas2022b} comes with two main limitations that are rooted in the usage of the distributions $F[\phi]$ and $G [\pi]$ as these describe field and momentum field configurations \textit{separately}. First, this REUR is restricted to bosonic modes and can not be straightforwardly extended to fermionic degrees of freedom, which excludes more than two-thirds of the elementary particles described by the Standard Model. Second, it does not capture the full information available in the field-theoretic phase space, since possible correlations between the field and the momentum field sectors are not described by $F[\phi]$ and $G[\pi]$. As a consequence, the usefulness of their associated (relative) entropies for studying the phenomena typically investigated in phase space, such as the thermalization of isolated quantum systems \cite{Deutsch2018} and quantum phase transitions (QPTs) \cite{Vojta2003}, is debatable. However, a quantitative understanding of the constraints imposed by the uncertainty principle could shed light on these and other statistical aspects of field theories. Further, the relation in \cite{Haas2022b} lacks invariance under rotations in phase space, which is a desirable property for detecting squeezing in arbitrary directions \cite{Wineland1994,Ma2011}.

To overcome these limitations, we put forward a field-theoretic generalization of the Husimi $Q$-distribution, which is a full phase-space representation of the quantum state defined as the diagonal elements in the coherent-state basis $Q [\phi, \pi] = \bTr \{\boldsymbol{\rho} \boldsymbol{\ket{\alpha} \bra{\alpha}} \}$ \cite{Husimi1940}. Its definition only relies on the notion of coherent states, which have been established for all physically relevant systems based on group-theoretic arguments \cite{Cahill1969,Radcliffe1971,Gilmore1974,Klauder1985,Zhang1990,Cahill1999,Combescure2012} and path integrals \cite{Ohnuki1978,Kamenev2011,Shankar2017}. Further, the Husimi $Q$-distribution bears the same experimental accessibility as $F[\phi]$ and $G[\pi]$, since it is the outcome distribution of the heterodyne detection protocol \cite{Schleich2001}. Albeit carrying the same information as the state itself, its experimental detection is substantially less demanding than quantum state tomography. It has been successfully reconstructed on a large variety of platforms, including photonic \cite{Noh1991,Noh1992,Landon2018}, cavity QED \cite{Kirchmair2013,Wang2016}, trapped ion \cite{Leibfried1996,Gaerttner2017} and ultracold atom \cite{Kunkel2019,Kunkel2021} systems as well as atomic gases in optical cavities \cite{Haas2014,Barontini2015}. 

Importantly, the Husimi $Q$-distribution is inherently non-negative (albeit being a quasi-probability distribution) \cite{Cartwright1976}, which enables the definition of the so-called Wehrl entropy $S[Q]$ \cite{Wehrl1978,Wehrl1979}. This entropy was originally introduced to interpolate between the Boltzmann entropy in classical phase space and the von Neumann entropy $S(\boldsymbol{\rho})$. As such, it naturally encodes the uncertainty principle via a state-independent lower bound $S[Q] \ge S[\bar{Q}] \sim N$ known as the Wehrl-Lieb inequality \cite{Lieb1978}, showing that quantum phase-space distributions can \textit{not} be localized to phase-space volumes $\le \hbar$. This implies the Wehrl entropy $S[Q]$ to be positive, which is generally not true for $S[F]$ and $S[G]$, and hence the Wehrl entropy can safely be regarded as a \textit{faithful} measure of the (quantum) information contained in phase space and as an entropy in the context of statistical physics and thermodynamics. 

The Wehrl-Lieb inequality has been generalized to various quantum systems \cite{Wehrl1979,Lieb1978,Carlen1991,Schupp1999,Luo2000,Lieb2014b,Lieb2016,Lieb2021,Schupp2022,Kulikov2022}, and recently also to fermions \cite{Haas2024}, thereby demonstrating the generality of the phase-space approach. Hence, the Husimi $Q$-distribution and the Wehrl (relative) entropy are the perfect tools to state the uncertainty principle for quantum fields in most general terms and to study its implications.

\subsection{Our contribution}
We employ the concept of entropic distinguishability for the field-theoretic Husimi $Q$-distribution to formulate the uncertainty principle in \textit{phase space} for bosonic as well as fermionic quantum fields. After setting up the Husimi $Q$-distributions and its entropy for field theories in \autoref{sec:QuantumFieldInPhaseSpace}, we derive our main result, the general REUR given in Eq. \eqref{eq:REUR} and its form for free fields \eqref{eq:REURFreeFields}, in \autoref{sec:REUR}. We apply these relations to the bosonic scalar field and the fermionic description of the transverse-field Ising model in \autoref{sec:ScalarField} and \autoref{sec:IsingModel}, respectively. In both cases, we prove that the entropic distinguishability of arbitrary states with respect to the vacuum is bounded by the total number of excitations. Additionally, we investigate how the uncertainty relation is intertwined with relativity, demonstrate the usefulness of the REUR for detecting squeezing for the scalar field, and use the REUR to unveil the quantum phase transition in the Ising model. Detailed computations and thorough discussions concerning the continuum and infinite volume limits on the lattice are provided in \hyperref[app:ScalarField]{Appendix A.1} and \hyperref[app:IsingModel]{Appendix A.2} for bosonic and fermionic fields, respectively.

\subsection*{Notation}
We use natural units $\hbar = k_{\text{B}} = 1$, write quantum operators and corresponding traces with bold letters, e.g., $\boldsymbol{\rho}$ and $\bTr \{ \boldsymbol{\rho} \}$, respectively, and denote vacuum expressions by a bar, e.g., $\bar{Q}$. Inner products in momentum space are compactly written as $\boldsymbol{\phi} \cdot \boldsymbol{\phi}$, both for lattice and continuous theories.

\section{Quantum fields in phase space}
\label{sec:QuantumFieldInPhaseSpace}

We consider a general quantum field operator $\boldsymbol{\phi} (x)$ together with its conjugate momentum $\boldsymbol{\pi} (x)$ on the real line $x \in \mathbb{R}$ in the Schrödinger picture \cite{Hatfield2018}. Statistical and algebraic properties of $\boldsymbol{\phi} (x)$ and $\boldsymbol{\pi} (x)$ are encoded in (anti-)commutation relations, which we keep open at this point. Since a large class of Hamiltonians $\boldsymbol{H}$ becomes diagonal after a Fourier transform, we work in momentum space in the remainder.

\subsection{Husimi $Q$-distribution}
Phase-space distributions such as the Husimi $Q$-distribution rely on coherent states. We construct coherent states starting from creation and annihilation operators $\boldsymbol{a}^{\dagger} (p)$ and $\boldsymbol{a} (p)$, respectively, which act as ladder operators with respect to the Hamiltonian's ground state $\ket{0}$, defined via $\boldsymbol{a} (p) \ket{0} = 0$ for all momenta $p$. The mode operators are generally linearly related to the fundamental fields in the sense that $\boldsymbol{a}(p) \sim \boldsymbol{\phi} (p) + i \boldsymbol{\pi} (p)$ and hence inherit their algebraic properties from the (anti-)commutation relations of $\boldsymbol{\phi} (p)$ and $\boldsymbol{\pi} (p)$. We introduce general field-theoretic coherent states as displaced vacuum states \cite{Zhang1990,Cahill1999}
\begin{equation}
    \ket{\alpha} \equiv \boldsymbol{D} [\chi] \ket{0},
    \label{eq:CoherentStatesDefinition}
\end{equation}
where the unitary displacement operator generating translations in phase space is defined as
\begin{equation}
    \boldsymbol{D} [\chi] = e^{\boldsymbol{a}^{\dagger} \cdot \alpha - \alpha^* \cdot \boldsymbol{a}}.
    \label{eq:DisplacementOperatorDefinition}
\end{equation}
The parameter $\alpha (p)$ is the complex-valued phase field, which we write in terms of the classical fields $\phi (p), \pi (p)$ in complete analogy to the annihilation operator $\alpha (p) \sim \phi (p) + i \pi (p)$, formally grouped into a single field $\chi = (\phi, \pi)^T$ covering phase space. For bosonic and fermionic degrees of freedom, one may equally define coherent states \eqref{eq:CoherentStatesDefinition} as eigenstates of the annihilation operator
\begin{equation}
    \boldsymbol{a} (p) \ket{\alpha} = \alpha (p) \ket{\alpha},
\end{equation}
(note that this definition is not always reasonable, consider, e.g., the $su(2)$ algebra). 

Importantly, coherent states resolve the identity in the sense of a functional integral 
\begin{equation}
    \mathds{1} = \int \mathcal{D} \chi \, \boldsymbol{\ket{\alpha} \bra{\alpha}},
\end{equation}
with the functional integral measure $\mathcal{D} \chi \sim \mathcal{D}\phi \mathcal{D} \pi$ being rigorously defined on the lattice, see \hyperref[app:ScalarFieldSetup]{Appendix A.1} and \hyperref[app:IsingModelSetup]{Appendix B.1} for bosonic and fermionic degrees of freedom, respectively. Consequently, coherent state projectors constitute a positive operator-valued measure (POVM) with measurement outcomes distributed according to the functional Husimi $Q$-distribution
\begin{equation}
    Q [\chi] = \bTr \{\boldsymbol{\rho} \boldsymbol{\ket{\alpha} \bra{\alpha}} \}.
    \label{eq:HusimiQDistributionDefinition}
\end{equation}
This distribution furnishes field-theoretic phase space, i.e., associates a functional quasi-probability to all classical field configurations for every quantum state $\boldsymbol{\rho}$. As such, it bears the same information content as the state itself.

\subsection{Absolute and relative Wehrl entropies}
From an information-theoretic point of view, the Husimi $Q$-distribution has two desirable properties: it is bounded (in particular non-negative), i.e., $0 \le Q [\chi] \le 1$, and normalized to unity with respect to the phase-space integral measure $\int \mathcal{D} \chi \, Q [\chi] = \bTr \{ \boldsymbol{\rho} \} = 1$. These properties enable the definition of the Wehrl entropy \cite{Wehrl1978,Wehrl1979}
\begin{equation}
    S [Q] = - \int \mathcal{D} \chi \, Q [\chi] \, \ln Q [\chi],
\end{equation}
which behaves as a classical entropy in the sense of being monotonic under partial trace. It converges to the Boltzmann entropy, more precisely, to the $H$-entropy appearing in the $H$-theorem, in the classical limit $\hbar \to 0$ and is an upper bound to the von Neumann entropy $S [\boldsymbol{\rho}] \le S [Q]$, corroborating its interpretation of a coarse-grained quantum entropy in phase space \cite{Wehrl1978,Wehrl1979}.

The Wehrl entropy is constrained by the uncertainty principle in the form of the Wehrl-Lieb inequality \cite{Wehrl1979,Lieb1978,Carlen1991,Schupp1999,Luo2000,Lieb2014b,Lieb2016,Lieb2021,Schupp2022,Kulikov2022,Haas2024}
\begin{equation}
    S [Q] \ge S [\bar{Q}] \sim N,
    \label{eq:WehrlLiebInequality}
\end{equation}
where $\bar{Q}$ is the Husimi $Q$-distribution of the decoupled vacuum. At this point, a few remarks are in order. First, the classical limit $\hbar \to 0$ implies $S [\bar{Q}] \to - \infty$, showing that classical phase-space distributions can be arbitrarily localized, explaining the possibility of the Boltzmann entropy attaining negative values. Second, the Wehrl-Lieb inequality contains information about correlations between $\phi(p)$ and $\pi(p)$ leading to both sides of \eqref{eq:WehrlLiebInequality} being invariant under rotations in phase space. As a result, the Wehrl-Lieb inequality is tighter than the EUR underlying the REUR presented in \cite{Haas2022b} almost everywhere \cite{Haas2021b}, which renders derived entanglement witnesses particularly strong \cite{Haas2022a,Haas2022c,Haas2022d}. Third, the inequality \eqref{eq:WehrlLiebInequality} is tight if and only if $Q$ corresponds to a coherent state. Hence, in contrast to many of the most well-known uncertainty relations, squeezed states exhibit an increment in Wehrl entropy. Fourth, no other type of entropy expresses the uncertainty principle in such a general way. Namely, by an inequality valid for continuous variables \cite{Wehrl1978,Lieb1978,Lieb2014b}, spin systems \cite{Lieb1978,Lieb2014b}, spinor Bose-Einstein-condensates and the strong interaction \cite{Lieb2016}, hyperbolic spacetimes \cite{Lieb2021, Kulikov2022} and fermions \cite{Haas2024}. Fifth, the Wehrl-Lieb inequality shows that the Wehrl entropy -- just as every entropy over some measurement distribution -- scales with the number of modes and thus diverges in both the continuum as well as the infinite volume limits. The divergence of such \textit{absolute} entropies is similar to the infrared divergence of the absolute energy in quantum field theories, which is conveniently cured by considering energy differences with respect to the vacuum instead.

Therefore, we introduce the Wehrl relative entropy between a distribution of interest $Q$ and some reference distribution $\tilde{Q}$ in a functional sense by adapting the standard definition for finite systems \cite{Kullback1951,Kullback1959,Cover2006}
\begin{equation}
     S [Q \| \tilde{Q}] =\int \mathcal{D} \chi \, Q[\chi] \left( \ln Q[\chi]- \ln \tilde{Q}[\chi] \right),
     \label{eq:WehrlRelativeEntropyDefinition}
\end{equation}
such that $S [Q \| \tilde{Q}] = 0$ if and only if $Q = \tilde{Q}$. It serves as a measure for the entropic distinguishability of $Q$ with respect to $\tilde{Q}$, but should not be confused with a true metric on the space of functional distributions, as it is not symmetric. The well-known invariance of the relative entropy for the transition from discrete to continuous variables \cite{Jaynes1968} carries over to the continuum and infinite volume limits. In this sense, \textit{entropic distinguishability} can be regarded as more universal than \textit{absolute entropy}, which motivates its usage for formulating uncertainty relations valid for quantum field theories.

\section{Relative entropic uncertainty relation (REUR)} 
\label{sec:REUR}

\subsection{Decomposition of the Wehrl relative entropy}

The Wehrl relative entropy \eqref{eq:WehrlRelativeEntropyDefinition} can be decomposed into differences of Wehrl entropies when choosing a reference distribution $\tilde{Q}=Q_{\text{max}}$ that extremizes the Wehrl entropy $S [\tilde{Q}]$ with respect to a given set of constraints $\xi [Q, \tilde{Q}]=$ const. \cite{Haas2021a}, which can be seen as follows. We start from writing (\ref{eq:WehrlRelativeEntropyDefinition}) with some reference $\tilde{Q}$ as
\begin{equation}
    S [Q \| \tilde{Q}] = - S [Q] + S [Q, \tilde{Q}],
\end{equation}
where
\begin{equation}
    S [Q, \tilde{Q}] = -\int \mathcal{D} \chi \, Q[\chi] \ln \tilde{Q}[\chi]
    \label{eq:WehrlCrossEntropyDefinition}
\end{equation}
denotes the Wehrl cross entropy (which also diverges in the field theory limit). Now, given a set of constraints
\begin{equation}
    \xi [ Q, \tilde{Q} ] = 0,
    \label{eq:Constraints}
\end{equation}
for example, equal covariance matrices, we want to choose the reference $\tilde{Q}$ such that the Wehrl relative entropy becomes a difference of entropies, namely 
\begin{equation}
    S [Q, \tilde{Q}] = S [\tilde{Q}],
\end{equation}
with the conditions \eqref{eq:Constraints} understood. As the latter relation has to hold for all $Q$, the reference distribution $\tilde{Q}$ has to extremize the entropy with respect to the constraints \eqref{eq:Constraints} and hence indeed $\tilde{Q} = Q_{\text{max}}$.

When generalizing this setting to the case
\begin{equation}
    \xi [Q, Q_\text{max}] = \kappa \neq 0,
    \label{eq:Constraints2}
\end{equation}
for example distributions with fixed but \textit{unequal} covariance matrices, the Wehrl relative entropy can not be fully reduced to an entropy difference. However, $Q_\text{max}$ is still of the functional form of the extremal entropy distribution with respect to the constraints \eqref{eq:Constraints2}, but is less optimal in the sense that its constrained parameters do not agree with those of the true distribution $Q$. As a result, the cross entropy acquires an additional term $\Xi [Q, Q_\text{max}]$ involving the constrained quantities appearing in $\xi [Q, Q_\text{max}]$, leading to the general decomposition of the Wehrl relative entropy
\begin{equation}
    S [Q \| Q_\text{max}] = S [Q_\text{max}] - S [Q] + \Xi [Q, Q_\text{max}].
    \label{eq:WehrlRelativeEntropyDecomposition}
\end{equation}

As an illustrative example, we consider the expectation value of some observable $\mathcal{O}$, which we assume to be fixed but different for $Q$ and $Q_\text{max}$, resulting in
\begin{equation}
    \begin{split}
        \xi [Q, Q_\text{max}] &= \int \mathcal{D} \chi \left( Q [\chi] - Q_\text{max} [\chi] \right) \mathcal{O} [\chi] \\
        &= \kappa,
    \end{split}
\end{equation}
with $\kappa \neq 0$. The corresponding extremal entropy distribution is of Boltzmann-type 
\begin{equation}
    Q_\text{max} [\chi] = \frac{1}{Z} e^{- \lambda \mathcal{O} [\chi]},
\end{equation}
with a Lagrange parameter $\lambda$ and a normalization constant $Z$. Then, we indeed find for the Wehrl cross entropy the linear decomposition
\begin{equation}
    \begin{split}
        &S [Q, Q_\text{max}] \\
        &= \ln Z + \lambda \int \mathcal{D} \chi \, Q [\chi] \mathcal{O} [\chi] \\ 
        &= \ln Z + \lambda \left( \kappa + \int \mathcal{D} \chi \, Q_{\text{max}} [\chi] \mathcal{O} [\chi] \right) \\
        &= S [Q_\text{max}] + \Xi [Q, Q_\text{max}],
    \end{split}
\end{equation}
with $\Xi [Q, Q_\text{max}] = \lambda \xi [Q, Q_\text{max}] \neq 0$. In particular, the second term is absent if and only if the expectation values with respect to $Q$ and $Q_\text{max}$ agree.

\subsection{General REUR}
A reformulation of the Wehrl-Lieb inequality \eqref{eq:WehrlLiebInequality} suitable for quantum fields is achieved by using the decomposition of the Wehrl relative entropy \eqref{eq:WehrlRelativeEntropyDecomposition}. Then, we immediately arrive at the general REUR in phase space
\begin{equation}
    S [Q \| Q_{\text{max}}] \le \Delta S [Q_{\max}, \bar{Q}] + \Xi [Q, Q_{\text{max}}],
    \label{eq:REUR}
\end{equation}
which poses an upper bound on the entropic distinguishability. Therein, we introduced the entropy difference with respect to the vacuum $\Delta S [Q_{\max}, \bar{Q}] = S [Q_{\text{max}}] - S [\bar{Q}] \ge 0$. 

Intuitively speaking, the REUR \eqref{eq:REUR} encodes the uncertainty principle by stating that some Husimi $Q$-distribution $Q$ can \textit{not} be distinguished arbitrarily well from a maximum entropy distribution $Q_{\text{max}}$. In the classical limit $\hbar \to 0$, the upper bound diverges since the vacuum distribution $\bar{Q}$ can be arbitrarily localized in phase space, resulting in $S[\bar{Q}] \to - \infty$. This shows that the upper bound in \eqref{eq:REUR} is strictly stronger than any classical bound, and thus of quantum origin. Importantly, all quantities appearing in \eqref{eq:REUR} remain invariant under the continuum and the infinite volume limits and are typically finite. Therefore, we claim that \eqref{eq:REUR} faithfully describes entropic uncertainty in most general terms by abiding by the notion of \textit{entropic distinguishability}. 

\subsection{Free quantum fields} 
The constraints $\xi$ and the reference distribution $Q_{\text{max}}$ can be chosen depending on the application. For free theories, Gaussian Husimi $Q$-distributions are of particular importance, which are of the form
\begin{equation}
    Q [\chi] =\frac{1}{Z} \, e^{-\frac{1}{2} \chi^T \cdot C^{-1} \cdot \chi},
    \label{eq:GaussianHusimiQ}
\end{equation}
where $Z$ is a normalization constant and
\begin{equation}
    C (p,p') = \int \mathcal{D} \chi \, Q [\chi] \, \chi (p) \chi (p')
\end{equation}
denotes the (unconnected) covariance matrix of the Husimi $Q$-distribution. Such distributions constitute maximum entropy distributions for a fixed covariance $C$. 

Since the vacuum $\bar{Q}$ is also Gaussian for free theories and additionally known to minimize uncertainty relations for finite systems, it serves as a natural reference distribution for entropic distinguishability -- analogous to the energy. This choice results in $\Delta S [\bar{Q}, \bar{Q}] = 0$ (see \autoref{subsec:SecondMomentURs} for a different choice). In order to determine the functional $\Xi [Q, \bar{Q}]$, we first evaluate the Wehrl entropy of the vacuum
\begin{equation}
    \begin{split}
        S [\bar{Q}] &= \ln \bar{Z} + \frac{1}{2} \int \mathcal{D} \chi \, \bar{Q} [\chi] \, \chi^T \cdot \bar{C}^{-1} \cdot \chi \\
        &= \ln \bar{Z} + \frac{1}{2} \Tr \{ \bar{C}^{-1} \bar{C}^T \}.
    \end{split}
\end{equation}
Analogously, we compute the Wehrl cross entropy of some arbitrary Husimi $Q$-distribution with covariance matrix $C$ with respect to the vacuum
\begin{equation}
    \begin{split}
        \hspace{-0.4cm}S [Q, \bar{Q}] &= - \int \mathcal{D} \chi \, Q [\chi] \, \ln \bar{Q} [\chi] \\
        &= \ln \bar{Z} + \frac{1}{2} \int \mathcal{D} \chi \, Q [\chi] \, \chi^T \cdot \bar{C}^{-1} \cdot \chi \\
        &= S [\bar{Q}] + \frac{1}{2} \Tr \{ \bar{C}^{-1} \cdot (C^T - \bar{C}^T) \},
    \end{split}
\end{equation}
which shows that
\begin{equation}
    \begin{split}
        \Xi [Q, \bar{Q}] = \frac{1}{2} \Tr \{ \bar{C}^{-1} \cdot(C^T - \bar{C}^T) \}.
    \end{split}
\end{equation}
Then, the general REUR \eqref{eq:REUR} implies the REUR for free quantum fields
\begin{equation}
    S [Q \| \bar{Q}] \le B = \frac{1}{2} \Tr \left\{ \bar{C}^{-1} \cdot \left(C^T - \bar{C}^T \right) \right\}.
    \label{eq:REURFreeFields}
\end{equation}
Remarkably, the bound $B$ contains only second-order correlation functions and thus can be computed for arbitrary states, especially when a calculation of the relative entropy itself is infeasible. Further, the finiteness of the bound becomes apparent through the difference term $C^T-\bar{C}^T$, which has \textit{finitely} many non-zero entries for any finite number of excitations, even in the continuum. We support this claim by evaluating the REUR \eqref{eq:REURFreeFields} for bosonic and fermionic quantum fields in the following.

\subsection{Second-moment uncertainty relations}
\label{subsec:SecondMomentURs}
When considered for a single mode, EURs imply their second-moment-based counterparts, as Gaussian distributions maximize entropies for fixed variances -- the most well-known example being Heisenberg's relation $\sigma_x \sigma_p \ge 1/2$ following from the relation of Białynicki-Birula and Mycielski $S(f) + S(g) \ge \ln (e \pi)$ \cite{Bialynicki-Birula1975,Coles2017,Hertz2019}. In the following, we show that also in this regard starting from a relative-entropic description is favorable as it additionally allows for a quantification of non-Gaussianity. 

We consider the reference distribution to be the Gaussian distribution with the same covariance matrix as the state of interest $C = C_{\text{max}}$, such that $\Xi [Q, Q_{\text{max}}] = 0$  and
\begin{equation}
    \Delta S [Q_{\text{max}}, \bar{Q}] = \frac{1}{2} \ln \frac{\det C}{\det \bar{C}}.
\end{equation}
Using the latter two equations in \eqref{eq:REUR} results in
\begin{equation}
    \frac{\det C}{\det \bar{C}} \ge e^{2 S [Q \| Q_\text{max}]} \ge 1,
    \label{eq:SecondMomentUR}
\end{equation}
which resembles the Robertson-Schrödinger uncertainty relation \cite{Robertson1930,Schroedinger1930} up to the fact that $C$ is the covariance of the Husimi $Q$-distribution and not the Wigner $W$-distribution (note also that both determinants diverge in the continuum while their fraction is finite). Eq. \eqref{eq:SecondMomentUR} nicely demonstrates that the REUR \eqref{eq:REUR} is stronger than its second-moment variant precisely when $Q$ is non-Gaussian, with the non-Gaussianity being measured by the Wehrl relative entropy $S[Q \| Q_\text{max}]$ between $Q$ and its closest Gaussian version $Q_\text{max}$.

\section{Scalar field} 
\label{sec:ScalarField}

\subsection{Husimi $Q$-distributions and covariance matrices}
We start with a relativistic scalar field theory for a real bosonic quantum field $\boldsymbol{\phi} (x)$ of mass $m$ and its conjugate momentum field $\boldsymbol{\pi} (x) = \partial_t \boldsymbol{\phi} (x)$. Their canonical commutation relation $[\boldsymbol{\phi} (x), \boldsymbol{\pi} (x')] = i \delta (x-x')$ forms a representation of the Heisenberg-Weyl algebra over the real numbers. The real-space Hamiltonian 
\begin{equation}
    \boldsymbol{H} = \frac{1}{2} \int \mathrm{d}x \left[ \boldsymbol{\pi}^2 + (\partial_x \boldsymbol{\phi})^2 + m^2 \boldsymbol{\phi}^2 \right]
    \label{eq:ScalarFieldHamiltonian}
\end{equation}
becomes diagonal in momentum space, where it is characterized by the relativistic dispersion relation $\omega^2 (p) = m^2 + p^2$ (see \hyperref[app:ScalarFieldSetup]{Appendix A.1}).

The Husimi $Q$-distribution of the ground state follows from $\braket{0 | \alpha} = \exp (-\alpha^* \cdot \alpha / 2)$, where $\alpha (p) = \left[ \omega (p) \phi (p) + i \pi (p) \right]/\sqrt{2 \omega (p)}$ is a complex-valued c-number for bosonic systems, and reads
\begin{equation}
    \bar{Q} [\chi] = e^{- \frac{1}{2} \left( \omega \phi \cdot \phi + \frac{1}{\omega} \pi \cdot \pi \right)},
    \label{eq:BosonicHusimiQVacuum}
\end{equation}
with unit normalization with respect to the measure $\mathcal{D} \chi = \mathcal{D}\phi \mathcal{D}\pi/(2 \pi)$ understood, see \hyperref[app:ScalarFieldVacuum]{Appendix A.2.1}. It corresponds to a Gaussian distribution centered around the origin in phase space with the covariance matrix $\bar{C} = \text{diag} (\bar{C}^{\phi}, \bar{C}^{\pi})$ being diagonal in a continuous sense, namely
\begin{equation}
    \begin{split}
        \bar{C}^{\phi} (p,p') &= \frac{2 \pi}{\omega (p)} \delta (p-p'), \\
        \bar{C}^{\pi} (p,p') &= 2 \pi \omega (p) \delta (p-p').
    \end{split}
    \label{eq:BosonicCovarianceMatrixVacuum}
\end{equation}
We refer to \hyperref[app:ScalarFieldFundamentalCorrelators]{Appendix A.2.2} for how $\bar{C}$ is related to the fundamental two-point correlators.

Eq. \eqref{eq:BosonicCovarianceMatrixVacuum} implies that the bound on the entropic distinguishability with respect to the vacuum only depends on the diagonal elements of the covariance matrix $C$ of the state of interest. For a general (possibly entangled) state $\boldsymbol{\rho}$, the corresponding Husimi $Q$-distribution acquires involved polynomial corrections to its vacuum form (see \hyperref[app:ScalarFieldExcitedStates]{Appendix A.2.3}). While the off-diagonal elements of the covariance matrix (which is symmetric $C=C^T$ in bosonic theories) are of similar complexity, the diagonal elements, which have to be understood in a functional sense since $C^{\phi, \pi} (p,p) \sim \delta (0)$, attain the simple form
\begin{equation}
    C^{\phi, \pi} (p,p) = \bar{C}^{\phi, \pi} (p,p) \left[ 1 + \braket{n (p)} \pm T (p) \right].
    \label{eq:BosonicCovarianceExcited}
\end{equation}
Therein, $\braket{n (p)} = \bTr \{ \boldsymbol{\rho} \, \boldsymbol{a}^{\dagger} (p) \boldsymbol{a} (p) \} \in \mathbb{R}^+_0$ denotes the average number of particles per mode $p$ and $T (p)$ involves a sum over off-diagonal contributions in the Fock basis, whose expression is provided in \eqref{eq:BosonicTl}. This shows that bosonic excitations manifest themselves in \textit{additive} contributions to the corresponding vacuum entries modulo the $T (p)$ terms.

\subsection{Bosonic REUR}
\label{subsec:BosonicREUR}
When computing the bound on the entropic distinguishability of an arbitrary state \eqref{eq:REURFreeFields}, we note that the two $T (p)$ terms appearing in $C^{\phi} (p,p)$ and $C^{\pi} (p,p)$ come with opposite signs and therefore cancel each other in the bound when evaluating the trace. First, we consider the case of finitely many excited momentum modes, for which the relative energy $E - \bar{E}$ remains finite. Importantly, the differences on the diagonal $C^{\phi, \pi} (p,p) - \bar{C}^{\phi, \pi} (p,p)$ are non-zero only for the finite number of excited modes, and hence all other contributions in the trace vanish. This leads to
\begin{equation}
    S [Q \| \bar{Q}] \le B = \sum_{p} \braket{n (p)},
    \label{eq:BosonicREUR}
\end{equation}
which shows that the entropic distinguishability of arbitrary states is constrained by nothing but the total particle number. We stress that this result holds in the continuous theory, since all possible divergencies appearing when taking the field theory limit canceled out.

The REUR \eqref{eq:BosonicREUR} describes, for instance, finitely many quasi-particles in a finite set of excited modes, in which case the particle number is definite in every mode $\braket{n (p)} = n (p)$. In contrast, when all momentum modes are occupied, the total particle number, and thus also the bound, diverge in the infinite volume limit. This is, for instance, the case for thermal excitations following the Bose-Einstein distribution $\braket{n (p)} = 1/(e^{\omega (p)/T}-1)$, where $T$ denotes the temperature. This behavior is again analogous to the relative energy of the system, where $E - \bar{E} \sim L$ for a thermal state when $T>0$, which results from the theory being defined on an interval of infinite length $L \sim \delta (0)$. Thus, a description in terms of an entropic distinguishability \textit{density} should be preferred, where the infinite volume factor is divided out on both sides of the REUR \eqref{eq:BosonicREUR}, resulting in the divergence-free relation
\begin{equation}
    s [Q \| \bar{Q} ] \le b = \int \mathrm{d}p \braket{n (p)}.
    \label{eq:BosonicREURDensity}
\end{equation}

\begin{figure*}[t]
    \centering
    \includegraphics[width=0.98\textwidth]{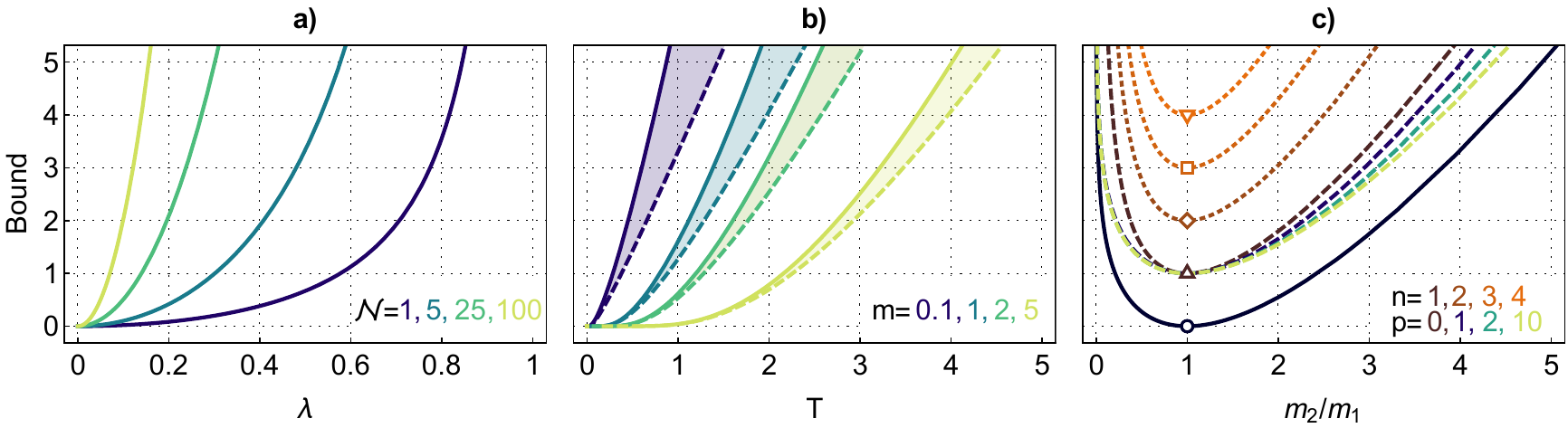}
    \caption{Bounds on the entropic distinguishability for a bosonic scalar field. \textbf{a)} $\mathcal{N}$ particle pairs in one- or two-mode squeezed vacuum states, see \eqref{eq:OMSV} and \eqref{eq:TMSV}, respectively, as a function of the squeezing parameter $\lambda$. \textbf{b)} Relativistic (straight) and non-relativistic (dashed) thermal states as a function of the temperature $T$ for several masses $m$. \textbf{c)} States of masses different from the underlying vacuum as a function of the mass ratio $m_2/m_1$: vacuum (straight), various particle numbers $n$ (dotted), and several momenta $p$ (dashed).}
    \label{fig:ScalarField}
\end{figure*}

\subsection{Applications}
\label{subsec:BosonicApplications}
Let us now analyze the bound on the entropic distinguishability for three physically relevant scenarios. First, we ask how squeezing (possibly with entanglement among excited modes) influences the bound. To that end, we consider the archetypal state describing squeezing and entanglement between two quasi-particles of opposite momenta $p$ and $-p$, that is, the two-mode squeezed vacuum (TMSV) 
\begin{equation}
    \sqrt{1- \lambda^2} \sum_{n (p) =0}^{\infty} \lambda^{n (p)} \ket{n (p)} \otimes \ket{n (-p)},
    \label{eq:TMSV}
\end{equation}
where $\lambda \in [0,1]$ denotes the squeezing parameter ($\lambda = 0$ corresponds to two uncorrelated vacua). Besides its importance in continuous-variable quantum information \cite{Weedbrook2012,Serafini2017}, the TMSV appears in field-theoretic scenarios when particle pairs are being generated as a result of a change in the total energy (note here that the single particles need to have opposite momenta by momentum conservation). This includes, for instance, quantum fields on curved backgrounds \cite{Hawking1975,Birrell1982,Mukhanov2007,Haas2022e,Haas2022f,Haas2022g} and quenches \cite{Chen2021}. Its uncorrelated counterpart is a tensor product over two one-mode squeezed vacuum (OMSV) states
\begin{equation}
    \sqrt[4]{1- \lambda^2} \sum_{n(p)=0}^{\infty} \frac{\sqrt{(2n (p))!}}{2^{n (p)} n (p)!}\lambda^{n (p)} \ket{2 n (p)},
    \label{eq:OMSV}
\end{equation}
with momenta $p$ and $-p$. For both states, we find the particle numbers in the excited modes $p$ and $-p$ to be $\braket{n (\pm p)} = \frac{\lambda^2}{1- \lambda^2}$. Hence, the bound for the entropic distinguishability of $\mathcal{N}$ equally squeezed particle pairs with respect to the vacuum is independent of their momenta and their entanglement and reads
\begin{equation}
    B = \frac{2 \mathcal{N} \lambda^2}{1- \lambda^2}.
    \label{eq:BosonicSqueezing}
\end{equation}
We plot the bound for $\mathcal{N}=1,5,25,100$ squeezed particle pairs in \autoref{fig:ScalarField} \textbf{a)}, showing its monotonic increase with the squeezing parameter $\lambda$. Note that $B \to 0$ for $\lambda \to 0$ (corresponding to two uncorrelated vacua), and $B \to \infty$ for $\lambda \to 1$, where the divergence grows with the number of squeezed particles, i.e., the total amount of squeezing. Since both sides of the REUR are invariant under rotations in phase space, the Wehrl relative entropy is a faithful and accessible measure for squeezing along arbitrary axes in the case of pure state, which can be advantageous over standard methods \cite{Wineland1994,Ma2011} when optimizing over squeezing directions becomes costly in terms of experimental runs. In a simple routine, squeezing of a -- possibly mixed -- state can be certified by measuring the covariance matrices of the momentum modes $p$ and $-p$, diagonalizing these matrices via rotations in phase space, and checking whether at least one entry on the diagonals is smaller than one, i.e., below the so-called shot-noise limit. For pure states, the amount of squeezing is then quantified by the Wehrl relative entropy, with the bound \eqref{eq:BosonicSqueezing} constraining the maximum amount of squeezing allowed by the uncertainty principle.

Second, we ask in which way relativity enters our uncertainty relation. Although the relativistic dispersion relation canceled out in deriving \eqref{eq:BosonicREUR}, average particle numbers may still implicitly depend on it. This is the case, for instance, for the thermal state via the Bose-Einstein distribution. Here, we compare the relativistic theory, where $\omega (p) = \sqrt{m^2 + p^2}$, to the non-relativistic theory, which is characterized by $\omega (p) = m + \frac{p^2}{2m}$, i.e., the leading order contribution to the relativistic expression for $p \ll m$. While the integral in the bound \eqref{eq:BosonicREURDensity} can only be evaluated numerically in the relativistic case, a closed analytical formula exists for the non-relativistic case, namely, $b=\sqrt{2 \pi m T} \, \text{Li}_{1/2} (e^{-m/T})$, where $\text{Li}_{a} (x)$ is the polylogarithm. We compare the two bounds as functions of the temperature for varying masses in \autoref{fig:ScalarField} \textbf{b)}. Independent of temperature and mass, we find the relativistic bounds (straight curves) to be larger than their non-relativistic counterparts (dashed curves). This observation is a consequence of the Bose-Einstein distribution being monotonically decreasing with $\omega(p)$ and relativistic excitations being less dispersive than non-relativistic ones, i.e., $\sqrt{m^2 + p^2} \le m + \frac{p^2}{2m}$ for all $m>0$ and $p \in \mathbb{R}$, which is an instance of Bernoulli's inequality. Hence, the entropic distinguishability of thermal fields is more strongly constrained for non-relativistic theories than for relativistic theories.

Third, we ask how well vacua and quasi-particle excitations of distinct masses can be distinguished -- a question for which a relative formulation of the uncertainty principle is particularly predestined. Let $m_1$ be the mass of the vacuum reference and $m_2$ be the mass of the state of interest. Then, the corresponding dispersion relations $\omega_1 (p)$ and $\omega_2 (p)$ do \textit{not} cancel when computing the bound \eqref{eq:BosonicREURDensity}. For particles, i.e., definite $\braket{n(p)} = n (p)$, we find
\begin{equation}
    \begin{split}
        b &= \int \mathrm{d} p \left[ \frac{1 \hspace{-0.05cm} + \hspace{-0.05cm}\braket{n (p)}}{2} \left(\frac{\omega_{1}(p)}{\omega_{2} (p)} \hspace{-0.05cm} + \hspace{-0.05cm} \frac{\omega_{2} (p)}{\omega_{1} (p)}\right) \hspace{-0.05cm} - \hspace{-0.05cm} 1 \right] \\
        &= m_1 \hspace{-0.05cm} \left(1 \hspace{-0.05cm} + \hspace{-0.05cm} \frac{m_2^2}{m_1^2} \right) \hspace{-0.05cm} K \hspace{-0.05cm} \left(1 \hspace{-0.05cm} - \hspace{-0.05cm}\frac{m_2^2}{m_1^2} \right) \hspace{-0.05cm} - \hspace{-0.05cm} 2 m_2 E \hspace{-0.05cm} \left( 1 \hspace{-0.05cm} - \hspace{-0.05cm} \frac{m_1^2}{m_2^2} \hspace{-0.05cm} \right) \\
        &\hspace{0.4cm}+ \frac{1}{2} \sum_p n (p) \left[\frac{\omega_{1}(p)}{\omega_{2} (p)} + \frac{\omega_{2} (p)}{\omega_{1} (p)}\right],
    \end{split}
    \label{eq:BosonicREURDistinctVacua}
\end{equation}
where $K(x)$ and $E(x)$ denote the complete elliptic integrals of the first and second kind, respectively. This reveals that the vacuum of mass $m_2$ may be interpreted as a highly excited state with respect to the vacuum of mass $m_1$. We show the result in \autoref{fig:ScalarField} \textbf{c)} as a function of the mass ratio $m_2/m_1$ with $m_1 = 1$ for two vacua (straight curve), $n=1,2,3,4$ particles at rest $p=0$ (dotted curves) and $n=1$ particle with momenta $p=0,1,2,10$ (dashed curves). Despite the appearance of elliptic integrals, we observe the bound \eqref{eq:BosonicREURDistinctVacua} to be finite for $m_2/m_1 \in (0, \infty)$. It attains its unique minimum value (note that the bound is convex), which is given by the total particle number $\sum_p n (p)$, precisely at $m_2=m_1$, i.e., when the two underlying vacua are identical. Further, the entropic distinguishability is generally more constrained for fast particles, since the mass difference is less relevant when $p \gg m_1, m_2$. Remarkably, the bound diverges logarithmically in the massless limit $m_2 \to 0$ in all cases (see \hyperref[app:ScalarFieldDistinctVacua]{Appendix A.3} for a discussion of the asymptotic behavior near $m_2 /m_1 \to 0, 1, \infty$), revealing that the REUR \eqref{eq:BosonicREURDensity} does \textit{not} restrict critical excitations. This indicates a quantitative connection between the uncertainty principle and critical phenomena, which we will analyze in more detail in \autoref{subsec:FermionicApplications}.

\section{Transverse-field Ising model}
\label{sec:IsingModel}

\subsection{Husimi $Q$-distributions and covariance matrices}
We now consider the transverse-field Ising model for a chain of spin-(1/2) operators $\{\boldsymbol{\sigma}^x_j, \boldsymbol{\sigma}^y_j, \boldsymbol{\sigma}_j^z \}$ forming a $su(2)$ algebra $[\boldsymbol{\sigma}^n_j, \boldsymbol{\sigma}^m_{j'}] = 2 i \epsilon_{n m o} \boldsymbol{\sigma}_j^o \delta_{j j'}$, described by the Hamiltonian 
\begin{equation}
    \boldsymbol{H} = - \sum_j \left( J \boldsymbol{\sigma}_j^z \boldsymbol{\sigma}_{j+1}^z + h \boldsymbol{\sigma}_j^x \right),
    \label{eq:IsingModelHamiltonian}
\end{equation}
where $J$ denotes the nearest-neighbor coupling and $h$ specifies the interaction strength of the spins with the external magnetic field. It is well-known that the discretized spinless Majorana fermion arises from this model after applying a Jordan-Wigner transformation \cite{Jordan1928,Lieb1961}. This maps the spin-(1/2) operators to fermionic operators via $\boldsymbol{\sigma}_j^x = 1 - 2 \epsilon \boldsymbol{\psi}^{\dagger}_j \boldsymbol{\psi}_j$ and $\boldsymbol{\sigma}_z^j = - \sqrt{\epsilon} \prod_{j'=1}^{j-1} \boldsymbol{\sigma}_{j'}^x (\boldsymbol{\psi}_j + \boldsymbol{\psi}^{\dagger}_{j})$, where $\epsilon > 0$ denotes a lattice spacing. The complex-valued fermionic operators are mode operators and as such obey anti-commutation relations $\{\boldsymbol{\psi}_{j}, \boldsymbol{\psi}^{\dagger}_{j'} \} = (1/\epsilon) \delta_{j j'}$, thereby constituting a representation of the Clifford algebra. Taking the continuum and infinite volume limits leads to the spinless Majorana fermion on the real line 
\begin{equation}
    \boldsymbol{H} = - \int \mathrm{d} x  \left[ \frac{v}{2} \boldsymbol{\psi}^{\dagger} \partial_x \boldsymbol{\psi} + \text{c.c.} + \gamma \abs{\boldsymbol{\psi}}^2 \right],
    \label{eq:MajoranaFermionHamiltonian}
\end{equation}
where $v = 2 \epsilon J$ and $ \gamma=2 (J-h)$. After applying a Bogoliubov transformation, its diagonal form is obtained in momentum space with the dispersion relation $\omega^2(p) = \gamma^2 + v^2 p^2$, which reads $\omega_l^2 = 4 \left[ J^2 + h^2 - 2 J h \cos \left( \epsilon \Delta k l \right) \right]$ for a finite Ising chain, see \hyperref[app:IsingModelSetup]{Appendix B.1}.

The construction of fermionic coherent states via \eqref{eq:CoherentStatesDefinition} requires the phase field $\alpha (p) = \sqrt{\omega (p) /2} \left[ \phi (p) + i \pi (p) \right]$ (as well as the real-valued fields $\phi$ and $\pi$) to be Grassmann-valued. As a consequence, phase-space integrals have to be understood as Berezin integrals with measure $\mathcal{D} \chi = \mathcal{D}\pi\mathcal{D}\phi\, i/\omega$. Analogously to the bosonic case, the vacuum Husimi $Q$-distribution 
\begin{equation}
    \bar{Q} [\chi] = e^{- \frac{i \omega}{2} \left(\pi\cdot\phi-\phi\cdot\pi\right) }
\end{equation}
is centered around the origin $\braket{\chi}=0$, which is implied by the simple properties of Berezin integrals. Note here that \textit{all} physical fermionic states need to have vanishing field expectation values as any non-vanishing field expectation value would be proportional to a Grassmann number, see \cite{Friis2013,Friis2016,Hackl2021}. The covariance matrix for fermions is anti-symmetric $C^T = - C$ and for the vacuum characterized by vanishing diagonal blocks
\begin{equation}
    \bar{C} = 
    \begin{pmatrix}
    0 & \bar{C}^{\phi\pi} \\
    \bar{C}^{\pi\phi} & 0
    \end{pmatrix},
    \label{eq:FermionicCovarianceMatrixVacuum}
\end{equation}
with 
\begin{equation}
    \bar{C}^{\phi\pi}(p,p') = - \bar{C}^{\pi\phi}(p,p') = \frac{i}{w(p)}\delta (p - p'),
    \label{eq:FermionicCovarianceMatrixVacuum2}
\end{equation} 
see \hyperref[app:FermionsVacuum]{Appendix B.2.1}. Hence, the bound \eqref{eq:REURFreeFields} only depends on the off-diagonal blocks of the covariance matrix $C$ of the state of interest, which are generally characterized by the diagonal elements (see \hyperref[app:IsingModelExcitedStates]{Appendix B.2.2})
\begin{equation}
    C^{\phi\pi}(p,p) = - C^{\pi\phi}(p,p) = \bar{C}^{\phi\pi}(p,p) \left(1 - \braket{n (p)} \right).
    \label{eq:FermionicCovarianceMatrixExcitations}
\end{equation}
In striking contrast to the bosonic case, cf. Eq. \eqref{eq:BosonicCovarianceExcited}, fermionic excitations appear as \textit{subtractive} contributions. As a result of Pauli's principle, the particle numbers are now restricted to the unit interval $\braket{n (p)} \in [0,1]$ and the $T (p)$ terms vanish.

\subsection{Fermionic REUR}
When evaluating the bound on the entropic distinguishability of excitations \eqref{eq:REURFreeFields}, the aforementioned minus sign in front of $\braket{n(p)}$ is canceled by the anti-symmetry of the fermionic covariance matrix. This leads to the very same bounds as in the bosonic case
\begin{equation}
    \begin{split}
        S [Q \| \bar{Q}] &\le B = \sum_{p} \braket{n (p)}, \\ 
        s [Q \| \bar{Q}] &\le b = \int \mathrm{d}p \, \braket{n (p)},
    \end{split}
    \label{eq:FermionicREUR}
\end{equation}
for finitely and infinitely many excited modes, respectively. Therefore, the entropic distinguishability of free bosonic and fermionic fields is constrained in precisely the same way, namely, by the total particle number.

\subsection{Applications}
\label{subsec:FermionicApplications}
Let us first comment on the thermal state. Analogous to the bosonic case, we immediately find the bound \eqref{eq:FermionicREUR} to be smaller in the non-relativistic limit than in the relativistic theory, since the Fermi-Dirac distribution $\braket{n (p)} = 1/(e^{\omega(p)/T} - 1)$ is monotonically decreasing with the dispersion $\omega (p)$, just as the Bose-Einstein distribution.

As we will now show, the fermionic REUR \eqref{eq:FermionicREUR} is particularly useful to study the distinguishability of phases and the appearance of quantum phase transitions (QPTs). While classical phase transitions are caused by thermal fluctuations present at finite temperatures, QPTs are driven by quantum fluctuations occurring at zero temperature, which arise as a result of the uncertainty principle. Consequently, many measures probing QPTs are based on an uncertainty relation \cite{Wineland1994,Ma2011}. However, a complete, and in particular quantitative, understanding of the relation between QPTs and the uncertainty principle beyond heuristic arguments remained elusive.

\begin{figure}[t]
    \centering
    \includegraphics[width=0.98\columnwidth]{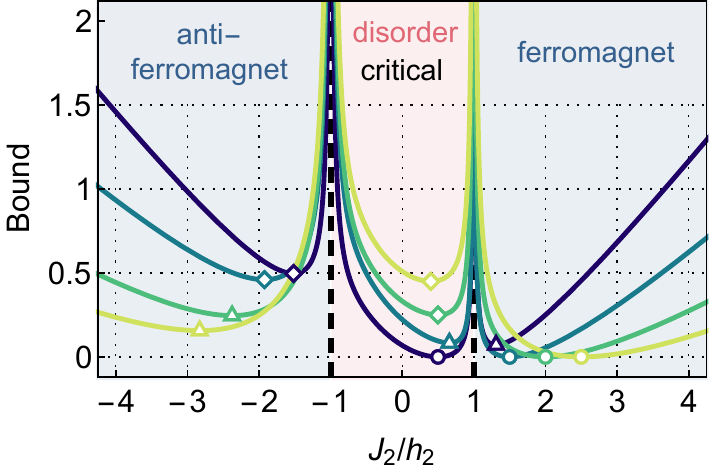}
    \caption{Bound on the entropic distinguishability for the spinless Majorana fermion as a function of $J_2/h_2$ with respect to the vacua of the disordered phase $J_1/h_1 = 0.5$ (dark blue) and the ordered phase $J_1/h_1=1.5$ (petrol), $J_1/h_1=2$ (green), $J_1/h_1=2.5$ (yellow). All curves diverge at the quantum phase transition $\abs{J_2/h_2}=1$ and attain their local minima (ascending order: circle, triangle, diamond) in different phases depending on the ratio $J_1/h_1$.}
    \label{fig:IsingModel}
\end{figure}

To that end, we consider a reference vacuum characterized by the fixed couplings $J_1$ and $h_1 = 1$, while we take the state of interest to be the vacuum with variable $J_2$ and $h_2 = 1$. This resembles the analysis underlying \autoref{fig:ScalarField} \textbf{c)} and results in the bound depending on the corresponding dispersion relations, see \eqref{eq:BosonicREURDistinctVacua} with $n(p)=0$ and $\gamma=2(J-h)$ playing the role of a mass $m$. We show the resulting bound on the entropic distinguishability for lattice spacing $\epsilon = 1$ and $N=10$ spins in \autoref{fig:IsingModel} for various choices for the reference $J_1/h_1$ covering the ordered (petrol, green and yellow) and the disordered (dark blue) phases (when the reference is at the critical point $\abs{J_1/h_1}=1$, the bound diverges for all $\abs{J_2/h_2} \neq 1$). Remarkably, all curves diverge precisely at the QPT $\abs{J_2 / h_2} = 1$ (black dashed line), thereby signaling critical behavior of second order, which shows that the entropic distinguishability is \textit{unconstrained} by the uncertainty principle at the critical point. Note here that in any classical description, the bound would evaluate to zero (not to infinity), as there is only one ground state. Operationally, an infinite relative entropy implies that the two compared distributions are truly distinguishable in the sense that the distribution of interest assigns positive probability densities to field configurations that are not contained in the model description. Hence, it is indeed the uncertainty principle that allows the ground state to become distinguishable enough around the QPT, which enables a complete rearrangement of its order beyond. 

Here, we stress that divergencies at the QPT are by no means rare -- the covariance matrix itself diverges at $p \to 0$ since the critical theory is gapless. However, when considering any absolute formulation of the uncertainty principle, this only causes the chosen uncertainty measure to diverge while the (lower) bounds remain unaltered. Therefore, standard uncertainty relations can not immediately explain why the uncertainty principle is necessary for the occurrence of QPTs. For entropic uncertainty relations, the situation is even worse, since any absolute entropy diverges independent of the state under consideration. Thus, to single out physical divergences, like those associated with the QPT itself, relative entropies are required.

Further, we remark that the fermionic bound \eqref{eq:FermionicREUR} comes with two symmetries: exchanging $h_1 \leftrightarrow J_1$ leaves the curves invariant, while flipping any of the signs of $J_1, h_1, J_2, h_2$ results in the curves being mirrored around the $y$-axis. Keeping these symmetries in mind, we observe from \autoref{fig:IsingModel} a hierarchy of local minima depending on the phase of the vacuum reference. Of course, the global minimum is always attained if and only if $J_1 = J_2$ and $h_1 = h_2$. Note here that changing $h_2 \neq h_1$ shifts the phase transition(s) accordingly and causes the global minimum to be strictly positive. If the reference vacuum is in the disordered phase $J_1 / h_1 = 0.5$ (dark blue), the first local minimum (triangle) is attained around $J_2 / h_2 \approx 1.3$, i.e., in the ferromagnetic regime of the ordered phase, while the second minimum (diamond) lies in the anti-ferromagnetic regime around $J_2 / h_2 \approx -1.5$. For ferromagnetic reference vacua, the first minimum occurs in the disordered phase as long as $J_1 / h_1 \lessapprox 2$ (petrol) and in the anti-ferromagnetic phase when $J_1 / h_1 \gtrapprox 2$ (yellow). This trend persists towards the limit $J_1 / h_1 \to \infty$, i.e., when the external magnetic field of the reference is irrelevant compared to the spin-spin coupling, along which the bound becomes symmetric around the $y$-axis. The only other case where the bound exhibits this symmetry appears for the state of interest attaining the opposite limit $J_2 / h_2 \to 0$. All these findings can be summarized as follows. The least distinguishable states always lie in the same phase as the vacuum reference. If the vacuum reference is in the disordered phase, the next local minimum is attained in the ordered phase with the same sign as $J_1/h_1$. If it is in one of the two ordered phases instead, the next minimum lies in the disordered phase if $J_1/h_1 \lessapprox 2$ and in the oppositely ordered phase when $J_1 / h_1 \gtrapprox 2$.

\section{Discussion}
We have put forward the REUR in phase space, which bounds the entropic distinguishability of quantum field configurations with respect to a reference distribution from above, and evaluated it for free fields of bosonic and fermionic type. We found that for arbitrary states the bound reduces to the total particle number, independent of the number of modes and the algebraic properties of the field theory under consideration. Thus, and by abiding by the generality of coherent states, we consider the REUR in phase space a universal way of stating the uncertainty principle in terms of entropic measures. Besides its generality, we found the REUR to be useful for various applications, which culminated in its ability to explain the occurrence of QPTs as a result of the entropic distinguishability becoming unbounded.

Let us provide a few closing comments. First, we remark that although we have studied the Ising model via its fermionic dual, the same type of analysis can be carried out on the level of the spin operators. In this case, the Husimi $Q$-distribution is defined with respect to spin coherent states, the associated phase space is isomorphic to a sphere \cite{Zhang1990} and the corresponding Wehrl-Lieb inequality depends on the total spin \cite{Lieb2014b}. In this context, the resulting REUR could be useful for witnessing spin squeezing and for constraining entropic distinguishability \textit{locally}.

Second, it is of particular interest to find REURs in more involved scenarios. This includes field theories with non-Abelian symmetry groups where field configurations are degenerate as a result of gauge symmetries, as well as interacting theories, where the vacuum is not of Gaussian form anymore, presumably resulting in a more involved bound on $S[Q \| \bar{Q}]$. 

Third, we stress that our results hold not only in the continuum and within a lattice regularization, but also using any other regularization scheme. For instance, one may equally work with wave packages, i.e., distribution-valued operators integrated against finitely-peaked test functions such as narrow Gaussians, which is of particular relevance for experimental investigations constrained by finite resolution.

Fourth, while the present work entirely focuses on uncertainty in momentum space, one may equally consider the REUR \eqref{eq:REUR} in position space and ask whether it encodes the entanglement inherently present in field theories. Remarkably, as shown in \cite{Haas2024b,Haas2024c,Haas2024d}, the most characteristic feature of entanglement in field theories, that is, the ground-state area law of the entanglement entropy, is present also for classical entropies (including the Wehrl entropy), provided that the extensive contributions dictated by the uncertainty principle are carefully removed. These findings naturally complement our results and further support the general usefulness of relative phase-space entropies for studying quantum phenomena. 

\section*{Acknowledgements}  
We thank Markus Schröfl and Stefan Floerchinger for early discussions on the subject and Martin Gärttner, Serge Deside and Célia Griffet for valuable comments on earlier versions of the manuscript. S.D. was supported by the ERC Starting Grant 949279 HighPHun. T.H. acknowledges support from the European Union under project ShoQC within the ERA-NET Cofund in Quantum Technologies (QuantERA) program as well as from the F.R.S.-FNRS under project CHEQS within the Excellence of Science (EOS) program.


\onecolumn
\appendix

\section{Scalar field}
\label{app:ScalarField}

\subsection{Setup}
\label{app:ScalarFieldSetup}
For a thorough discussion of all issues related to the continuum and infinite volume limits, we introduce a lattice regularization $x \to \epsilon j, \boldsymbol{\phi} (x) \to \boldsymbol{\phi}_j$ with lattice spacing $\epsilon$ and restrict to a finite interval $L= N \epsilon$ with $N \in \mathbb{N}$ modes, which discretizes the momenta $p \to \Delta k l, \boldsymbol{\phi}(p) \to \boldsymbol{\phi}_l$ with spacing $\Delta k = 2 \pi/L$. Then, the continuum limit corresponds to $\epsilon \to 0, \, N \to \infty$ with $L = N \epsilon $ fixed, while in the infinite volume limit we take $\Delta k \to 0, \, N \to \infty$ with $\epsilon = L / N$ fixed. We refer to taking both limits as the field theory limit. 

Starting from the Hamiltonian \eqref{eq:ScalarFieldHamiltonian}, the regularization procedure leads to a Hamiltonian for finitely many modes $j \in \{1,...,N\}$ (we assume periodic boundary conditions $\boldsymbol{\phi}_{0} = \boldsymbol{\phi}_N$)
\begin{equation}
    \boldsymbol{H} = \frac{1}{2} \sum_j \epsilon \left[ \boldsymbol{\pi}_j^2+\frac{1}{\epsilon^2}(\boldsymbol{\phi}_j-\boldsymbol{\phi}_{j-1})^2+m^2\boldsymbol{\phi}_j^2 \right],
    \label{eq:ScalarHamiltonianDiscretized}
\end{equation}
and discretized commutation relations
\begin{equation}
    [\boldsymbol{\phi}_j, \boldsymbol{\pi}_{j'}] = \frac{i}{\epsilon} \delta_{j j'}, \quad [\boldsymbol{\phi}_j, \boldsymbol{\phi}_{j'}] = [\boldsymbol{\pi}_j, \boldsymbol{\pi}_{j'}] = 0.
\end{equation}
We perform a Fourier transformation to momentum space 
\begin{equation}
    \boldsymbol{\phi}_j=\sum_l \frac{\Delta k}{2 \pi} e^{i \Delta k l \epsilon j} \tilde{\boldsymbol{\phi}}_l,
    \label{eq:DFT}
\end{equation}
where $l \in \{-N/2+1, ..., N/2 \}$ (we assume $N$ to be even without loss of generality) denotes the discrete momentum modes, while $\tilde{\boldsymbol{\phi}}_l$ is complex-valued with $\tilde{\boldsymbol{\phi}}^{\dagger}_l = \tilde{\boldsymbol{\phi}}_{-l}$, and similarly for $\boldsymbol{\pi}_j$. Then, Kronecker-$\delta$'s in position and momentum space have the representations
\begin{equation}
    \frac{1}{\epsilon} \delta_{j j'} = \sum_l \frac{\Delta k}{2 \pi} e^{i \Delta k l \epsilon (j-j')}, \quad \frac{2 \pi}{\Delta k} \delta_{l l'} = \sum_j \epsilon \, e^{i \epsilon j \Delta k (l - l')},
\end{equation}
respectively, and the Hamiltonian \eqref{eq:ScalarHamiltonianDiscretized} becomes
\begin{equation}
    \boldsymbol{H} = \frac{1}{2} \sum_l \frac{\Delta k}{2 \pi} \left( \abs{\tilde{\boldsymbol{\pi}}_l}^2 + \omega_l^2 \abs{\tilde{\boldsymbol{\phi}}_l}^2 \right).
\end{equation}
We apply a unitary canonical transformation
\begin{equation}
    \tilde{\boldsymbol{\phi}}_l =\frac{1}{2}(1+i) \boldsymbol{\phi}_l+\frac{1}{2}(1-i) \boldsymbol{\phi}_{-l}, \quad \tilde{\boldsymbol{\pi}}_l =\frac{1}{2}(1-i) \boldsymbol{\pi}_l+\frac{1}{2}(1+i) \boldsymbol{\pi}_{-l},
    \label{eq: trafo dimensionless fields}
\end{equation}
to render the fields in momentum space real $\boldsymbol{\phi}_l,\boldsymbol{\pi}_l \in \mathbb{R}$, such that the discrete commutation relations of the momentum-space fields $\boldsymbol{\phi}_l$ and $\boldsymbol{\pi}_l$ read
\begin{equation}
    [\boldsymbol{\phi}_l,\boldsymbol{\pi}_{l'}] = i \frac{2\pi}{\Delta k} \delta_{l l'}, \quad [\boldsymbol{\phi}_l, \boldsymbol{\phi}_{l'}] = [\boldsymbol{\pi}_l, \boldsymbol{\pi}_{l'}] = 0.
    \label{eq:DiscreteCanonicalCommutationRelation}
\end{equation} 
We end up with the momentum-space Hamiltonian
\begin{equation}
    \boldsymbol{H} = \frac{1}{2} \sum_l \frac{\Delta k}{2 \pi} \left( \boldsymbol{\pi}^2_l + \omega_l^2 \boldsymbol{\phi}^2_l \right),
    \label{eq:ScalarHamiltonianMomentumSpace}
\end{equation}
with discrete frequencies
\begin{equation}
    \omega^2_l = \frac{4}{\epsilon^2}\sin^2 \left(\frac{\epsilon \Delta k l}{2} \right) + m^2.
\end{equation}
In the field theory limit we get back
\begin{equation}
    \boldsymbol{H} = \frac{1}{2} \int \frac{\mathrm{d}p}{2 \pi} \left[ \boldsymbol{\pi}^2(p) + \omega^2(p) \boldsymbol{\phi}^2(p) \right],
\end{equation}
with the standard relativistic dispersion relation
\begin{equation}
    \omega^2 (p) = p^2 + m^2.
\end{equation}
Now, we introduce creation and annihilation operators
\begin{equation}
    \boldsymbol{a}^{\dagger}_l = \frac{1}{\sqrt{2 \omega_l}} \left( \omega_l \boldsymbol{\phi}_l - i \boldsymbol{\pi} _l\right), \quad \boldsymbol{a}_l = \frac{1}{\sqrt{2 \omega_l}} \left( \omega_l \boldsymbol{\phi}_l + i \boldsymbol{\pi}_l \right),
    \label{eq:BosonicLadderOperators}
\end{equation}
respectively, which fulfill bosonic commutation relations
\begin{equation}
    [\boldsymbol{a}_l , \boldsymbol{a}_{l'}^\dagger] = \frac{2\pi}{\Delta k}\delta_{ll'}, \quad [\boldsymbol{a}_l , \boldsymbol{a}_{l'}] = [\boldsymbol{a}_l^\dagger , \boldsymbol{a}_{l'}^\dagger] = 0.
    \label{eq:BosonicCommutationRelationsAppendix}
\end{equation}
Expressed through the mode operators, the Hamiltonian \eqref{eq:ScalarHamiltonianMomentumSpace} takes the diagonal form
\begin{equation}
    \boldsymbol{H} = \sum_l \frac{\Delta k}{2\pi} \omega_l \left( \boldsymbol{a}_l^\dagger \boldsymbol{a}_l+ \frac{1}{2}\frac{2 \pi}{\Delta k} \right).
    \label{eq:ScalarHamiltonianMomentumSpaceDiagonal}
\end{equation}
In the second term, the dependence of the Hamiltonian on the number of modes $N$ becomes manifest. Taking either the continuum or the infinite volume limit, which both require $N \to\infty$, leads to a divergence since formally we have
\begin{equation}
    \boldsymbol{H} = \int \frac{\mathrm{d} p}{2\pi} \, \omega (p) \boldsymbol{a}^\dagger (p) \boldsymbol{a} (p) + \frac{1}{2} \int \mathrm{d} p \, \omega (p) \delta (0).
\end{equation}
Hence, the second term is usually omitted, which effectively corresponds to working with energy differences with respect to the vacuum instead of absolute energies in the continuous theory.

\subsection{Husimi $Q$-distributions}

\subsubsection{Vacuum}
\label{app:ScalarFieldVacuum}
We construct the vacuum Husimi $Q$-distribution $\bar{Q}$ from the displacement operator $\boldsymbol{D}[\chi]$, see Eq. \eqref{eq:DisplacementOperatorDefinition} in the main text, and its algebraic properties. We start from the commutator of the two operators appearing in the exponential
\begin{equation}
    \left[\boldsymbol{a}_l^{\dagger} \alpha_l, \alpha_{l'}^* \boldsymbol{a}_{l'} \right] = - \frac{2 \pi}{\Delta k} \alpha^*_{l'} \alpha_{l} \delta_{l l'},
\end{equation}
which evaluates to a real number. Thus, the Taylor series appearing in the Baker-Campbell-Hausdorff formula stops at first order, and we can write
\begin{equation}
    \boldsymbol{D} [\chi] = e^{\sum_l \frac{\Delta k}{2 \pi} \boldsymbol{a}_l^{\dagger} \alpha_l} e^{- \frac{1}{2} \sum_l \frac{\Delta k}{2 \pi} \alpha_l^* \alpha_l} e^{- \sum_l \frac{\Delta k}{2 \pi} \alpha^*_l \boldsymbol{a}_l}.
\end{equation}
In this form, it is easy to see that
\begin{equation}
        \braket{0 | \alpha} = \braket{0 | \boldsymbol{D} [\chi] | 0} = e^{- \frac{1}{2} \sum_l \frac{\Delta k}{2 \pi} \alpha_l^* \alpha_l} \braket{0 | \left(1 + \boldsymbol{a}^{\dagger}_l \alpha_l + \dots \right) \left(1 - \alpha^*_l \boldsymbol{a}_l \mp \dots \right) | 0} = e^{- \frac{1}{2} \sum_l \frac{\Delta k}{2 \pi} \alpha_l^* \alpha_l},
\end{equation}
leading to the vacuum Husimi $Q$-distribution
\begin{equation}
    \bar{Q} [\chi] = \abs{\braket{0 | \alpha}}^2 = e^{- \sum_l \frac{\Delta k}{2 \pi} \alpha^*_l \alpha_l}.
\end{equation}
The vacuum Husimi $Q$-distribution may also be written as
\begin{equation}
    \bar{Q} [\chi] = e^{- \frac{1}{2} \sum_l \frac{\Delta k}{2 \pi} \left( \omega_l \phi_l^2 + \frac{\pi^2_l}{\omega_l} \right)},
\end{equation}
which in the field theory limit becomes a functional quasi-probability density
\begin{equation}
    \bar{Q} [\chi] = e^{- \frac{1}{2} \int \frac{\mathrm{d}p}{2 \pi}\left( \omega (p) \phi^2 (p) + \frac{\pi^2 (p)}{\omega (p)} \right)}.
    \label{eq:BosonicHusimiQVacuumContinuum}
\end{equation}
For completeness, we check its normalization. Functional integrals over phase space are defined on the lattice as 
\begin{equation}
    \mathcal{D} \chi = \prod_l \frac{\mathrm{d}\phi_l \mathrm{d}\pi_l}{2 \pi} \frac{\Delta k}{2 \pi}.
\end{equation}
For practical purposes, we employ polar coordinates $(r_l, \varphi_l) \in [0, \infty) \times [0,2\pi)$ defined via
\begin{equation}
    \phi_l = \frac{r_l}{\sqrt{\omega_l}} \cos \varphi_l, \quad \pi_l = \sqrt{\omega_l} r_l \sin \varphi_l,
\end{equation}
with integral measure $\mathrm{d}\chi_l = (\mathrm{d}r_l \mathrm{d}\varphi_l)/(2 \pi) \, \Delta k/(2 \pi) \, r_l$ and $r_l = \sqrt{\omega_l \phi_l^2 + \pi_l^2/\omega_l}$. Then, we find
\begin{equation}
    \begin{split}
        \int \mathcal{D} \chi \, \bar{Q} [\chi] = \prod_{l} \int \frac{\mathrm{d}\phi_l \mathrm{d}\pi_l}{2 \pi} \frac{\Delta k}{2 \pi} \, e^{-\frac{1}{2} \frac{\Delta k}{2 \pi} \left( \omega_l \phi_l^2 + \frac{\pi^2_l}{\omega_l} \right)}  = \prod_l \int \frac{\mathrm{d}r_l \mathrm{d}\varphi_l}{2 \pi} \frac{\Delta k}{2 \pi} \, r_l \,  e^{- \frac{1}{2} \frac{\Delta k}{2 \pi} r_l^2} = 1.
    \end{split}
\end{equation}
The first- and second-order moments can be read off directly from \eqref{eq:BosonicHusimiQVacuum}, but we calculate them explicitly for illustrative purposes. The expectation values vanish since
\begin{equation}
    \begin{split}
        \braket{\overline{\phi_{l'}}} &= \int \mathcal{D} \chi \, \bar{Q} [\chi] \, \phi_{l'}\\
        &= \left[ \prod_{l \neq l'} \int \frac{\mathrm{d}\phi_l \mathrm{d}\pi_l}{2 \pi} \frac{\Delta k}{2 \pi} \, e^{-\frac{1}{2} \frac{\Delta k}{2 \pi} \left( \omega_l \phi_l^2 + \frac{\pi^2_l}{\omega_l} \right)} \right] \int \frac{\mathrm{d}\phi_{l'} \mathrm{d}\pi_{l'}}{2 \pi} \frac{\Delta k}{2 \pi} \, \phi_{l'} \, e^{-\frac{1}{2} \frac{\Delta k}{2 \pi} \left( \omega_{l'} \phi_{l'}^2 + \frac{\pi^2_{l'}}{\omega_{l'}} \right)} \\
        &= 1 \times \frac{1}{\sqrt{\omega_{l'}}} \int \mathrm{d}r_{l'} \frac{\Delta k}{2 \pi} \, r^2_{l'} \, e^{- \frac{1}{2} \frac{\Delta k}{2 \pi} r_{l'}^2} \int \frac{\mathrm{d}\varphi_{l'}}{2 \pi} \cos \varphi_{l'} \\
        &= 0,
    \end{split}
\end{equation}
and similarly $\braket{\overline{\pi_{l'}}}=0$, for all modes $l'$. Here, we used that for every mode $l \neq l'$, the corresponding distribution is normalized such that only the contribution from the mode $l'$ itself has to be considered. By the same reasoning, the off-diagonal blocks of the covariance matrix vanish $\bar{C}^{\phi \pi}_{l l'} = 0$. The only non-vanishing entries appear on the diagonal of the diagonal blocks, for which we obtain
\begin{equation}
    \bar{C}^{\phi}_{l l'} = \int \mathcal{D} \chi \, \bar{Q} [\chi] \, \phi_{l} \phi_{l'} = \frac{\delta_{l l'}}{\omega_{l}} \int \mathrm{d}r_{l} \frac{\Delta k}{2 \pi} \, r^3_{l} \, e^{- \frac{1}{2} \frac{\Delta k}{2 \pi} r_{l}^2} \int \frac{\mathrm{d}\varphi_{l}}{2 \pi} \cos^2 \varphi_l = \frac{2 \pi}{\Delta k} \frac{\delta_{l l'}}{\omega_l},
\end{equation}
and analogously,
\begin{equation}
    \bar{C}^{\pi}_{l l'} = \frac{2 \pi}{\Delta k} \omega_l \delta_{l l'},
\end{equation}
such that in total the vacuum Husimi $Q$-covariance matrix reads
\begin{equation}
    \bar{C} = 
    \begin{pmatrix}
        \bar{C}^{\phi} & 0 \\
        0 & \bar{C}^{\pi}

    \end{pmatrix} 
    = \frac{2 \pi}{\Delta k} 
    \begin{pmatrix}
        \frac{1}{\omega_l} \mathds{1} & 0 \\
        0 & \omega_l \mathds{1}
    \end{pmatrix}.
    \label{eq:BosonicVacuumCovariance}
\end{equation}
The continuous expressions are provided in Eq. \eqref{eq:BosonicCovarianceMatrixVacuum} in the main text. For its application in the bound of the REUR, we also give the inverse covariance matrices, which are found using the defining equations
\begin{equation}
    \sum_{l'} \frac{\Delta k}{2 \pi} \bar{C}_{l l'}^{-1} \bar{C}_{l' l''} = \frac{2 \pi}{\Delta k} \delta_{l l''}, \quad \int \frac{\mathrm{d}p'}{2\pi} \bar{C}^{-1}(p,p') \bar{C} (p',p'') = 2 \pi \delta (p-p''),
\end{equation}
for the lattice and the continuous theory, respectively, resulting in
\begin{equation}
    \bar{C}^{-1} = \frac{2 \pi}{\Delta k} \begin{pmatrix}
    \omega_l \mathds{1} & 0 \\
    0 & \frac{1}{\omega_l} \mathds{1}
    \end{pmatrix}.
    \label{eq:BosonicVacuumInverseCovariance}
\end{equation}

\subsubsection{Comment on relation to fundamental two-point correlation functions}
\label{app:ScalarFieldFundamentalCorrelators}
Let us here mention that \eqref{eq:BosonicVacuumCovariance} is related to the well-known vacuum two-point correlation functions of the free scalar field by a factor of two. More precisely, the Husimi $Q$-covariance obeys in general 
\begin{equation}
    C = \gamma + \bar{\gamma},
    \label{eq:CovarianceMatricesRelation}
\end{equation}
where $\gamma$ is the Wigner $W$-covariance matrix and, therefore, also the covariance matrix with respect to the density operator $\boldsymbol{\rho}$ by the Wigner-Weyl transformation. It is also of block form and contains the fundamental two-point correlation functions \cite{Weedbrook2012,Serafini2017}
\begin{equation}
    \gamma = \begin{pmatrix}
    \bTr \{ \boldsymbol{\rho} \, \boldsymbol{\phi} \boldsymbol{\phi} \} & \frac{1}{2} \bTr \{ \boldsymbol{\rho} \, \{ \boldsymbol{\phi}, \boldsymbol{\pi} \} \} \\
    \frac{1}{2} \bTr \{ \boldsymbol{\rho} \, \{ \boldsymbol{\phi}, \boldsymbol{\pi} \} \} & \bTr \{ \boldsymbol{\rho} \, \boldsymbol{\pi} \boldsymbol{\pi} \}
    \end{pmatrix}.
\end{equation}
Using the inverse of \eqref{eq:BosonicLadderOperators}, we find that for the vacuum the mixed two-point correlation functions vanish since $\braket{0 | \boldsymbol{\phi}_l \boldsymbol{\pi}_{l'} | 0} = - \braket{0 | \boldsymbol{\pi}_{l'} \boldsymbol{\phi}_{l} | 0} = \frac{i}{2} \delta_{l l'}$, while the homogeneous correlators evaluate to
\begin{equation}
   \braket{0 | \boldsymbol{\phi}_l \boldsymbol{\phi}_{l'} | 0} = \frac{2 \pi}{\Delta k} \frac{\delta_{l l'}}{2 \omega_l}, \quad \braket{0 | \boldsymbol{\pi}_l \boldsymbol{\pi}_{l'} | 0} = \frac{2 \pi}{\Delta k} \frac{\omega_l}{2} \delta_{l l'}.
\end{equation}
These expressions are in agreement with Eq. \eqref{eq:BosonicVacuumCovariance} with the additional factor $1/2$ following from \eqref{eq:CovarianceMatricesRelation}. In the field theory limit, we obtain the well-known correlators in momentum space $\braket{0 | \boldsymbol{\phi} (p) \boldsymbol{\pi} (p') | 0} = - \braket{0 | \boldsymbol{\pi} (p') \boldsymbol{\phi} (p) | 0} = \frac{i}{2} \delta (p-p')$ and
\begin{equation}
    \braket{0 | \boldsymbol{\phi} (p) \boldsymbol{\phi} (p') | 0} =  2 \pi \frac{\delta (p-p')}{2 \omega (p)}, \quad \braket{0 | \boldsymbol{\pi} (p) \boldsymbol{\pi} (p') | 0} = 2 \pi \frac{\omega (p)}{2} \delta (p-p').
\end{equation}

\subsubsection{Excited states}
\label{app:ScalarFieldExcitedStates}
We start from an excited state vector $\ket{n_l} \in \mathcal{H}$ with $n_l \in \mathbb{N}$ excitations in mode $l$, which we define as
\begin{equation}
    \ket{n_l} = \frac{1}{\sqrt{n_l !}} \left( \sqrt{\frac{\Delta k}{2 \pi}} \boldsymbol{a}_l^{\dagger} \right)^{n_l} \ket{0}.
    \label{eq:ExcitedStateScalarField}
\end{equation}
Therein, the factor $\sqrt{\Delta k / (2 \pi)}$ expresses the fact that excited states are non-normalizable in the field theory limit in the sense that the required normalization constant is formally proportional to $\delta (0)$. The states $\ket{n_l}$ span an orthonormal basis for the Hilbert space $\mathcal{H}$ in the sense that
\begin{equation}
    \braket{n_l | n'_{l'}} = \delta_{n n'} \delta_{l l'},
\end{equation}
showing that states corresponding to different modes or carrying differently many excitations in the same mode are orthogonal. Following this construction, the density operator for a general state can be expanded as
\begin{equation}
    \boldsymbol{\rho} = \sum_{\Vec{n}} \rho_{\Vec{n}} \bigotimes_l \boldsymbol{\ket{n_l} \bra{n'_l}},
    \label{eq:GeneralBosonicState}
\end{equation}
where $\Vec{n}=(n_{-N/2+1},n'_{-N/2+1},...,n_{N/2},n'_{N/2})$ contains indices $n_l$ and $n'_l$ for each mode $l$ and the sum $\sum_{\Vec{n}}=\sum_{n_{-N/2+1},n'_{-N/2+1},...,n_{N/2},n'_{N/2}=0}^{\infty}$ sum over all excitations $n_l$ and $n'_l$ for each mode $l$. This state is normalized by
\begin{equation}
    \sum_{\{\Vec{n}|n_l=n'_l \,\forall l\}} \rho_{\Vec{n}}= \sum_{\Vec{n}} \rho_{\Vec{n}}\,\prod_{l} \delta_{n'_{l} n_{l} } = 1.
\end{equation}  
We can describe states with a fixed as well as a fluctuating particle number by defining the total particle number per mode $l$
\begin{equation}
    \braket{n_l} = \bTr \{ \boldsymbol{\rho} \,  \boldsymbol{a}^{\dagger}_l \boldsymbol{a}_l \} = \sum_{\Vec{n}} \rho_{\Vec{n}}\,n_l \,\prod_{l'} \delta_{n'_{l'} n_{l'} }=\sum^{\infty}_{n_l=0} \rho^{(l)}_{n_ln_l}\,n_l \, ,
\end{equation}
where we identified the marginal density matrix of the $l$-th mode
\begin{equation}
    \rho^{(l)}_{n_ln'_l}=\Tr_{l'\neq l} \{\rho_{\Vec{n}} \}=\left(\prod_{l'\neq l} \sum_{n_{l'}, n'_{l'}}\delta_{n'_{l'} n_{l'} }\right)\rho_{\Vec{n}}.
\end{equation}
The corresponding Husimi $Q$-distribution follows by using the overlap formula
\begin{equation}
    \braket{n_l | \alpha} = \frac{1}{\sqrt{n_l !}} \left(\frac{\Delta k}{2 \pi}\right)^{n_l/2} \braket{0 | \boldsymbol{a}_l^{n_l} | \alpha} = \frac{1}{\sqrt{n_l !}} \left(\frac{\Delta k}{2 \pi}\right)^{n_l/2} \alpha_l^{n_l} \braket{0 | \alpha},
\end{equation}
leading to
\begin{equation}
    \begin{split}
        Q [\chi] &= \sum_{\Vec{n}} \rho_{\Vec{n}} \prod_l  \braket{\alpha | n_l} \braket{n'_l | \alpha} \\
        &= \sum_{\Vec{n}} \rho_{\Vec{n}} \prod_l  \frac{1}{\sqrt{n_l!n'_l!}} \left(\frac{\Delta k}{2 \pi} \frac{1}{2 \omega_l}\right)^{\frac{n_l+n'_l}{2}} \left(\omega_l \phi_l - i \pi_l\right)^{n_l} \left( \omega_l \phi_l + i \pi_l \right)^{n'_l} e^{- \frac{1}{2} \frac{\Delta k}{2 \pi} \left( \omega_l \phi_l^2 + \frac{\pi^2_l}{\omega_l} \right)}.
    \end{split}
\end{equation}
For an arbitrary state, the field expectation values are non-zero in general, but we omit their explicit form since they do not appear in the bound of the REUR for free fields \eqref{eq:REURFreeFields}. Note, however, that for the special case where the state is of product form and diagonal in the Fock basis, i.e., when $\rho_{n_l n'_l} \propto \delta_{n_l n'_l}$, we recover distributions centered around the origin with $\braket{\phi_l} = \braket{\pi_l} = 0$. For the covariance matrix, we note that all entries except the diagonal entries will be multiplied by zero when applied in the bound of the REUR. Hence, it is sufficient to consider
\begin{equation}
    \begin{split}
        C^{\phi}_{ll} &= \int \mathcal{D} \chi \, Q [\chi] \, \phi^2_{l} \\
        &=\sum_{\Vec{n}} \rho_{\Vec{n}} \left(\prod_{l'\neq l} \delta_{n'_{l'} n_{l'} }\right) \frac{1}{\sqrt{n_l!n'_l!}} \\&\hspace{4em}\times\int \mathrm{d}r_l \left(\frac{\Delta k}{2 \pi} \frac{1}{2}\right)^{\frac{n_l+n'_l}{2} + 1} 2 r^{3 + n_l + n_{l'}} e^{- \frac{1}{2} \frac{\Delta k}{2 \pi} r_l^2} \int \frac{\mathrm{d}\varphi_l}{2 \pi} \frac{1}{\omega_l} \cos^2 \varphi_l \, e^{i \varphi_l (n_l' - n_l)} \\
        &=\frac{2 \pi}{\Delta k} \,  \sum_{n_l,n'_l}  \frac{\rho^{(l)}_{n_ln'_l}}{\sqrt{n_l! n'_l!}} \, 2 \, \Gamma \left(\frac{4 + n_l + n'_l}{2} \right) \frac{2\delta_{n'_l n_l} + \delta_{n'_l (n_l + 2)} + \delta_{n'_l (n_l - 2)}}{4\omega_l} \\
        &= \frac{2 \pi}{\Delta k} \frac{1}{\omega_l} \left(1 + \braket{n_l} + T_l \right) \\
        &= \bar{C}_{ll}^{\phi} \left(1 + \braket{n_l} + T_l \right)
    \end{split}
    \label{eq:BosonicExcitedCovarianceField}
\end{equation}
and analogously 
\begin{equation}
    \begin{split}
        C_{ll}^{\pi} &= \bar{C}_{ll}^{\pi} \left(1 + \braket{n_l} - T_l \right),
    \end{split} 
    \label{eq:BosonicExcitedCovarianceMomentumField}
\end{equation}
where
\begin{equation}
    T_l = \frac{1}{2}\sum_{n_l}\left[\rho^{(l)}_{n_l (n_l+2)}\sqrt{(n_l + 2)(n_l + 1)} + \rho^{(l)}_{n_l (n_l-2)} \sqrt{n_l (n_l - 1)} \right].
    \label{eq:BosonicTl}
\end{equation}
Remarkably, for a diagonal Fock state, we have $T_l = 0$. Hence, free quasi-particle states as well as thermal states \textit{add} the mean particle number per mode to the vacuum covariance matrix. This also holds in the field theory limit, in which case both covariance matrices have to be understood as tempered distributions.

\subsection{Asymptotics of the bound for distinct vacua}
\label{app:ScalarFieldDistinctVacua}
\begin{figure}[h!]
    \centering
    \includegraphics[width=0.4\textwidth]{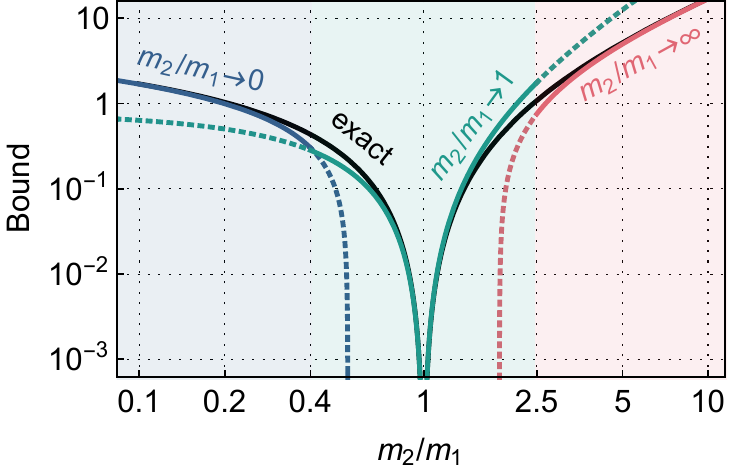}
    \caption{Double logarithmic plot of the bound (black) for two vacua as a function of the mass ratio distinct masses $m_2/m_1$ (we set $m_1=1$) together with its asymptotics in the limits $m_2/m_1 \to 0$ (blue), $m_2/m_1 \to 1$ (petrol) and $m_2/m_1 \to \infty$ (red).}
    \label{fig:ScalarFieldAppendix}
\end{figure}

We consider the first term in \eqref{eq:BosonicREURDistinctVacua}
\begin{equation}
    \bar{b} =  m_1 \left(1 + \frac{m_2^2}{m_1^2} \right) K \left(1-\frac{m_2^2}{m_1^2} \right) - 2 m_2 E \left( 1 - \frac{m_1^2}{m_2^2} \right),
\end{equation}
which sets a bound on the distinguishability of two vacua of distinct masses. Its complexity can be reduced when considering expansions for the three limits $m_2/m_1 \to 0, m_2/m_1 \to 1$, and $m_2/m_1 \to \infty$. To first order in $m_2/m_1$, we find the corresponding asymptotics
\begin{equation}
    \bar{b} \overset{m_2/m_1 \to 0}{=} m_1 \ln \left(\frac{4}{e^2} \frac{m_1}{m_2} \right), \quad \bar{b} \overset{m_2/m_1 \to 1}{=} \frac{\pi}{4} m_1 \left(1-\frac{m_2}{m_1} \right)^2, \quad \bar{b} \overset{m_2/m_1 \to \infty}{=} m_2 \ln \left(\frac{4}{e^2} \frac{m_2}{m_1} \right),
\end{equation}
which are compared to the exact result in \autoref{fig:ScalarFieldAppendix}. The bound diverges logarithmically in the massless limit (blue curve) while being quadratic to first order in $m_2/m_1$ around $m_2/m_1 = 1$ (petrol curve) as a result of convexity. For $m_2/m_1 \to \infty$ (red curve), the logarithmic divergence acquires an additional linear factor.

\section{Transverse-field Ising model}
\label{app:IsingModel}

\subsection{Setup}
\label{app:IsingModelSetup}
We construct the theory for a spinless Majorana fermion starting from the transverse-field Ising model with periodic boundary conditions \eqref{eq:IsingModelHamiltonian} (as before, the continuum and infinite volume limits are emphasized). When applying the Jordan-Wigner transformation (note that one sometimes finds an alternative convention in the literature, which is obtained for $\boldsymbol{\sigma}_j^x \to \boldsymbol{\sigma}_j^z$ and $\boldsymbol{\sigma}_j^z \to - \boldsymbol{\sigma}_j^x$)
\begin{equation}
    \boldsymbol{\sigma}_j^x = 1 - 2 \epsilon \boldsymbol{\psi}^{\dagger}_j \boldsymbol{\psi}_j, \quad \boldsymbol{\sigma}_z^j = - \sqrt{\epsilon} \prod_{j'=1}^{j-1} \boldsymbol{\sigma}_{j'}^x (\boldsymbol{\psi}_j + \boldsymbol{\psi}^{\dagger}_{j}),
\end{equation}
we obtain, after dropping the constant term, 
\begin{equation}
    \boldsymbol{H} = - \sum_j \epsilon \left[ J \left( \boldsymbol{\psi}_{j+1} \boldsymbol{\psi}_j + \boldsymbol{\psi}^{\dagger}_{j+1} \boldsymbol{\psi}_j + \boldsymbol{\psi}^{\dagger}_j \boldsymbol{\psi}_{j+1} + \boldsymbol{\psi}^{\dagger}_j \boldsymbol{\psi}^{\dagger}_{j+1} \right) - 2 h \boldsymbol{\psi}^{\dagger}_j \boldsymbol{\psi}_j \right].
    \label{eq:MajoranaHamiltonianDiscretized}
\end{equation}
The continuum limit $\epsilon \to 0, N \to \infty$ is achieved when approaching the phase transition at $h = J$ with $J \to \infty$ and $h/J \to 1$ for $v = 2 \epsilon J, \gamma=2 (J-h)$ fixed, while the infinite-volume limit requires $L \to \infty, N \to \infty$ for $\epsilon = L/N$ fixed as usual. Taking both limits results in the Hamiltonian for the spinless Majorana fermion on the real line given in Eq. \eqref{eq:MajoranaFermionHamiltonian}, for distribution-valued fermionic field operators fulfilling
\begin{equation}
    \{\boldsymbol{\psi} (x), \boldsymbol{\psi}^{\dagger} (x) \} = i \delta (x-x'), \quad \{\boldsymbol{\psi} (x), \boldsymbol{\psi} (x') \} = \{\boldsymbol{\psi}^{\dagger} (x), \boldsymbol{\psi}^{\dagger} (x') \} = 0.
    \label{eq:FermionicAntiCommutationRelations}
\end{equation}
We again employ a discrete Fourier transformation, where we assume periodic boundary conditions $\boldsymbol{\psi}_0 = \boldsymbol{\psi}_N$ and that the number of modes $N$ is even such that $l \in \{-N/2 + 1, ..., N/2 \}$, which results in
\begin{equation}
    \boldsymbol{\psi}_j = \sum_l \frac{\Delta k}{2 \pi} e^{i \Delta k l \epsilon j} \tilde{\boldsymbol{\psi}}_l.
    \label{eq:DFT2}
\end{equation}
Using the identities 
\begin{equation}
    \begin{split}
        \sum_j \epsilon \, \boldsymbol{\psi}_j^{\dagger} \boldsymbol{\psi}_j &= \sum_l \frac{\Delta k}{2 \pi} \tilde{\boldsymbol{\psi}^{\dagger}}_l \tilde{\boldsymbol{\psi}}_l, \\
        \sum_j \epsilon \, \boldsymbol{\psi}_{j+1} \boldsymbol{\psi}_j &= - i \sum_l \frac{\Delta k}{2 \pi} \sin \left( \Delta k \epsilon l \right) \tilde{\boldsymbol{\psi}}_{-l} \tilde{\boldsymbol{\psi}}_l, \\
        \sum_j \epsilon \, \boldsymbol{\psi}^{\dagger}_{j+1} \boldsymbol{\psi}_j &= \sum_l \frac{\Delta k}{2 \pi} e^{-i \Delta k \epsilon l} \tilde{\boldsymbol{\psi}}^{\dagger}_{l} \tilde{\boldsymbol{\psi}}_l, \\
        \sum_j \epsilon \, \boldsymbol{\psi}^{\dagger}_{j} \boldsymbol{\psi}_{j+1} &= \sum_l \frac{\Delta k}{2 \pi} e^{i \Delta k \epsilon l} \tilde{\boldsymbol{\psi}}^{\dagger}_{l} \tilde{\boldsymbol{\psi}}_l, \\
        \sum_j \epsilon \, \boldsymbol{\psi}^{\dagger}_{j} \boldsymbol{\psi}^{\dagger}_{j+1} &= - i \sum_l \frac{\Delta k}{2 \pi} \sin \left( \Delta k \epsilon l \right) \tilde{\boldsymbol{\psi}}^{\dagger}_{-l} \tilde{\boldsymbol{\psi}}^{\dagger}_l, \\
    \end{split}
\end{equation}
the position-space Hamiltonian \eqref{eq:MajoranaHamiltonianDiscretized} becomes in momentum space
\begin{equation}
    \boldsymbol{H} = \sum_l \frac{\Delta k}{2 \pi} \left\{ \kappa_l \tilde{\boldsymbol{\psi}}_l^{\dagger} \tilde{\boldsymbol{\psi}}_l + i \lambda_l \left( \tilde{\boldsymbol{\psi}}_{-l} \tilde{\boldsymbol{\psi}}_{l} + \tilde{\boldsymbol{\psi}}_{-l}^{\dagger} \tilde{\boldsymbol{\psi}}_{l}^{\dagger} \right) \right\},
    \label{eq:MajoranaHamiltonanMomentum}
\end{equation}
where we introduced the two real functions
\begin{equation}
    \kappa_l = 2 \left[ h - J \cos \left(\epsilon \Delta k l \right) \right], \quad \lambda_l = J \sin \left(\epsilon \Delta k l \right),
\end{equation}
for convenience. To diagonalize \eqref{eq:MajoranaHamiltonanMomentum}, we perform a Bogoliubov transformation on the fermionic mode operators
\begin{equation}
    \tilde{\boldsymbol{\psi}}_l = u_l \boldsymbol{\psi}_l + i v_l \boldsymbol{\psi}_{-l}^{\dagger},
    \label{eq:BogoliubovTransformation}
\end{equation}
with two real numbers $u_l, v_l$ fulfilling $u_{-l} = u_l, v_{-l} = - v_l$ as well as $u_l^2 + v_l^2 = 1$. The resulting momentum-space operators $\boldsymbol{\psi}_l$ serve as raising and lowering operators
\begin{equation}
    \boldsymbol{a}^{\dagger}_l = \boldsymbol{\psi}_l^{\dagger}, \quad \boldsymbol{a}_l = \boldsymbol{\psi}_l,
\end{equation}
and as such have to fulfill fermionic anti-commutation relations
\begin{equation}
    \{ \boldsymbol{\psi}_l, \boldsymbol{\psi}_{l'}^{\dagger} \} = (2\pi/\Delta k) \delta_{l l'}, \quad \{ \boldsymbol{\psi}_l, \boldsymbol{\psi}_{l'} \} = \{ \boldsymbol{\psi}_l^{\dagger}, \boldsymbol{\psi}_{l'}^{\dagger} \} = 0.
\end{equation}
Moreover, the phase field has to be Grassmann-valued with 
\begin{equation}
    \{ \alpha_l, \alpha^*_{l'} \} = \{ \alpha_l, \alpha_{l'} \} = \{ \alpha^{*}_l, \alpha^*_{l'} \} = 0
\end{equation}
 and additionally 
 \begin{equation}
     \{ \alpha_l, \boldsymbol{\psi}_{l'} \} = \{ \alpha_l, \boldsymbol{\psi}^{\dagger}_{l'} \} = \{ \alpha^{*}_l, \boldsymbol{\psi}_{l'} \} = \{ \alpha^{*}_l, \boldsymbol{\psi}^{\dagger}_{l'} \} = 0.
 \end{equation}
Then, we find the Hamiltonian
\begin{equation}
    \begin{split}
        \boldsymbol{H} = \sum_l \frac{\Delta k}{2 \pi} \Big\{& \left( \kappa_l u_l^2 + 2 \lambda_l u_l v_l \right) \boldsymbol{\psi}^{\dagger}_l \boldsymbol{\psi}_l + \left( - \kappa_l v_l^2 + 2 \lambda_l u_l v_l \right) \boldsymbol{\psi}^{\dagger}_{-l} \boldsymbol{\psi}_{-l} \\
        &- i \left[ \kappa_l u_l v_l - \lambda_l \left( u_l^2 - v_l^2 \right) \right] \left( \boldsymbol{\psi}_{-l} \boldsymbol{\psi}_l + \boldsymbol{\psi}^{\dagger}_{-l} \boldsymbol{\psi}^{\dagger}_l \right) \\
        &- \left( - \kappa_l v_l^2 + 2 \lambda_l u_l v_l \right) \frac{2 \pi}{\Delta k} \Big\}.
    \end{split}
    \label{eq:MajoranaHamiltonanMomentum2}
\end{equation}
The functions $u_l, v_l$ are chosen such that the off-diagonal term, i.e., the second line in the previous equation, vanishes
\begin{equation}
    0 = \kappa_l u_l v_l - \lambda_l \left( u_l^2 - v_l^2\right).
\end{equation}
This equation is conveniently solved by parameterizing $u_l, v_l$ in terms of trigonometric functions respecting their even/odd nature
\begin{equation}
    u_l = \cos \left(\frac{\vartheta_l}{2}\right), \quad v_l = \sin \left(\frac{\vartheta_l}{2}\right),
\end{equation}
for some angle $\vartheta_l \in [0, 2\pi)$, yielding
\begin{equation}
    \tan \vartheta_l = \frac{2 \lambda_l}{\kappa_l} = \frac{\sin \left(\epsilon \Delta k l \right) }{h/J - \cos \left(\epsilon \Delta k l \right)}.
\end{equation}
For this value of $\vartheta_l$, the remaining terms in \eqref{eq:MajoranaHamiltonanMomentum2} simplify to
\begin{equation}
    \kappa_l u_l^2 + 2 \lambda_l u_l v_l = \frac{\kappa_l + \omega_l}{2}, \quad -\kappa_l v_l^2 + 2 \lambda_l u_l v_l = \frac{-\kappa_l + \omega_l}{2},
\end{equation}
where we introduced the discrete frequencies
\begin{equation}
    \omega_l^2 = 4 \left[ J^2 + h^2 - 2 J h \cos \left( \epsilon \Delta k l \right) \right].
\end{equation}
Then, the fully diagonal momentum-space Hamiltonian becomes
\begin{equation}
    \boldsymbol{H} = \sum_l \frac{\Delta k}{2 \pi} \omega_l \left( \boldsymbol{\psi}_l^{\dagger} \boldsymbol{\psi}_l - \frac{1}{2} \frac{\Delta k}{2 \pi} \right),
\end{equation}
in close analogy to the diagonal scalar field Hamiltonian \eqref{eq:ScalarHamiltonianMomentumSpaceDiagonal}. In the field-theory limit, the latter reads
\begin{equation}
    \boldsymbol{H} = \int \frac{\mathrm{d} p}{2 \pi} \omega (p) \boldsymbol{\psi}^{\dagger} (p) \boldsymbol{\psi} (p) - \frac{1}{2} \int \mathrm{d} p \, \omega (p) \delta (0),
\end{equation}
where again, the infinite offset due to the vacuum energy appears. The corresponding relativistic dispersion relation is
\begin{equation}
    \omega^2(p) = \gamma^2 + v^2 p^2.
\end{equation}

\subsection{Husimi $Q$-distributions}

\subsubsection{Vacuum}
\label{app:FermionsVacuum}
We construct the vacuum Husimi $Q$-distribution $\bar{Q}$ similar to the bosonic case. Since
\begin{equation}
    \left[ \boldsymbol{\psi}_l^{\dagger} \alpha_l, \alpha_{l'}^{*} \boldsymbol{\psi}_{l'} \right] = - \frac{\Delta k}{2 \pi} \alpha_{l'}^{*} \alpha_l \delta_{l l'}
\end{equation}
commutes with $\boldsymbol{\psi}_l^{\dagger} \alpha_l$ and $\alpha_{l'}^{*} \boldsymbol{\psi}_{l'}$, the displacement operator decomposes according to the Baker-Campbell-Hausdorff formula as
\begin{equation}
    \boldsymbol{D} [\chi] = e^{\sum_l \frac{\Delta k}{2 \pi} \boldsymbol{\psi}^{\dagger}_l \alpha_l} \, e^{- \frac{1}{2} \sum_l \frac{\Delta k}{2 \pi} \alpha_l^* \alpha_l} \, e^{-\sum_l \frac{\Delta k}{2 \pi} \boldsymbol{\psi}_l \alpha^*_l}.
\end{equation}
This leads to the overlap
\begin{equation}
    \braket{0 | \alpha} = \braket{0 | \boldsymbol{D} [\chi] | 0} = e^{- \frac{1}{2} \sum_l \frac{\Delta k}{2 \pi} \alpha_l^* \alpha_l} = \prod_{l} \left( 1 + \frac{1}{2} \frac{\Delta k}{2 \pi} \alpha_l \alpha^*_l \right),
\end{equation}
and therefore the vacuum Husimi $Q$-distribution is given by (see \cite{Cahill1999} for the appearance of the additional minus sign in the definition of the fermionic Husimi $Q$-distribution)
\begin{equation}
    \bar{Q} [\chi] = \braket{\alpha | 0} \braket{0 | - \alpha} = e^{- \sum_l \frac{\Delta k}{2 \pi} \alpha_l^* \alpha_l} = \prod_{l} \left( 1 + \frac{\Delta k}{2 \pi} \alpha_l \alpha^*_l \right).
    \label{eq:FermionicHusimiQVacuum}
\end{equation}
Analogously to the bosonic case, we can decompose the complex phase field into two real Majorana fields according to
\begin{equation}
    \boldsymbol{a}_l = \sqrt{\frac{\omega_l}{2}} \left( \boldsymbol{\phi}_l + i \boldsymbol{\pi}_l \right).
\end{equation}
Just as the phase field, the classical fields $\phi_l$ and $\pi_l$ are now Grassmann valued, and their corresponding field operators satisfy anti-commutation relations 
\begin{equation}
    \{\boldsymbol{\phi}_l,\boldsymbol{\pi}_{l'}\} = 0, \quad \{\boldsymbol{\phi}_l, \boldsymbol{\phi}_{l'}\} = \{\boldsymbol{\pi}_l, \boldsymbol{\pi}_{l'}\} = \frac{i}{\omega_l} \frac{2\pi}{\Delta k} \delta_{l l'}.
    \label{eq:DiscreteCanonicalAntiCommutationRelation}
\end{equation}
In this basis, the discrete vacuum Husimi $Q$-distribution reads
\begin{equation}
    \bar{Q} [\chi] = \prod_{l} \left( 1 + \frac{\Delta k}{2 \pi}\frac{i \omega_l}{2} \left(\pi_l\phi_l-\phi_l\pi_l\right) \right).
    \label{eq:FermionicHusimiQVacuumRealFields}
\end{equation}
which becomes a functional quasi-probability density in the field theory limit.
We again check the normalization of the functional integral measure for completeness
\begin{equation}
    \int \mathcal{D} \chi \, \bar{Q}[\chi] = \int \prod_{l'} \left( \mathrm{d} \pi_{l'} \mathrm{d} \phi_{l'} \frac{2 \pi }{\Delta k} \frac{i}{\omega_l}\right) \prod_{l} \left( 1 + \frac{\Delta k}{2 \pi}\frac{i\omega_l}{2} \left(\pi_l\phi_l-\phi_l\pi_l\right) \right) = \prod_l \int \mathrm{d} \pi_l \mathrm{d} \phi_l \, \phi_{l} \pi_{l} = 1,
\end{equation}
where we have chosen the standard sign convention for Berezin integrals, that is, that the innermost integral is performed first. Note here that the fraction $(2\pi/\Delta k)$ in the integral measure is inverse to the corresponding term in the bosonic case, as a result of the Berezin integral being \textit{anti}-proportional to the Jacobian determinant after a variable transformation.

The vacuum field expectation values vanish $\braket{\overline{\chi_l}}=0$, which also holds for the diagonal blocks of the vacuum covariance matrix $\bar{C}^{\phi \phi}_{ll} = \bar{C}^{\pi \pi}_{ll} =0$, due to the Grassmanian nature of the considered variables. The entries of the off-diagonal block evaluate to
\begin{equation}
    \bar{C}^{\phi \pi}_{l l'} =\int \mathcal{D} \chi \, \bar{Q} [\chi] \, \phi_l \pi_{l'} = \int \mathrm{d} \pi_{l} \mathrm{d} \phi_{l} \frac{2 \pi }{\Delta k} \frac{i}{\omega_l} \left( 1 + \frac{\Delta k}{2 \pi}\frac{i\omega_l}{2} \left(\pi_l\phi_l-\phi_l\pi_l\right) \right) \phi_l \pi_{l} \delta_{l l'} = \frac{2\pi}{\Delta k} \frac{i}{\omega_l}\delta_{l l'},
\end{equation}
and by the anti-commutation relations
\begin{equation}
    \bar{C}^{\pi \phi}_{l l'} = - \bar{C}^{\phi \pi}_{l l'} =  -\frac{2\pi}{\Delta k} \frac{i}{\omega_l}\delta_{l l'},
\end{equation}
with the analogous results in the continuous theory given in Eq. \eqref{eq:FermionicCovarianceMatrixVacuum}. The structure of the vacuum covariance matrix is preserved under inversion, with the diagonal elements of the off-diagonal blocks acquiring a prefactor of $- \omega^2_l$.

\subsubsection{Excited state}
\label{app:IsingModelExcitedStates}
Similarly to the bosonic case, we define an excited state $\ket{n_l} \in \mathcal{H}$ with $n_l \in \{0,1\}$ excitations in mode $l$ by Pauli's principle as
\begin{equation}
    \ket{n_l} = \left( \sqrt{\frac{\Delta k}{2 \pi}} \boldsymbol{\psi}_l^{\dagger} \right)^{n_l} \ket{0}.
\end{equation}
Then, a general state reads
\begin{equation}
    \boldsymbol{\rho} = \sum_{\Vec{n}} \rho_{\Vec{n}} \bigotimes_l \boldsymbol{\ket{n_l} \bra{n'_l}},
\end{equation}
similarly to the bosonic state (\ref{eq:GeneralBosonicState}) with the difference, that the sums over expectations now only run from $0$ to $1$ $\sum_{\Vec{n}}=\sum_{n_{-N/2+1},n'_{-N/2+1},...,n_{N/2},n'_{N/2}=0}^{1}$. Computing the coherent state overlap simply gives
\begin{equation}
    \braket{n_l | \alpha} = \left( \frac{\Delta k}{2 \pi} \right)^{n_l/2} \alpha_l^{n_l} \braket{0 | \alpha},
\end{equation} 
and thus we find for the Husimi $Q$-distribution
\begin{equation}
    \begin{split}
        Q [\chi] &= \sum_{\Vec{n}} \rho_{\Vec{n}} \prod_l  \braket{\alpha | n_l} \braket{n'_l | - \alpha}\\
        &= \sum_{\Vec{n}} \rho_{\Vec{n}} \prod_l \left( \frac{\Delta k}{2 \pi} \right)^{\frac{n_l+n'_l}{2}} \alpha_l^{n'_l}\left(\alpha^*_l\right)^{n_l}   e^{- \frac{\Delta k}{2 \pi} \alpha^*_{l} \alpha_l}\\
        &= \sum_{\Vec{n}} \rho_{\Vec{n}} \prod_l \left( \frac{\Delta k}{2 \pi} \frac{\omega_l}{2 } \right)^{\frac{n_l+n'_l}{2}}  \left((i \pi_l)^{n'_l}\phi_l^{n_l}+\phi_l^{n'_l}(-i \pi_l)^{n_l}\right) \left( 1 + \frac{\Delta k}{2 \pi}\frac{i\omega_l}{2} \left(\pi_l\phi_l-\phi_l\pi_l\right)\right). 
    \end{split}
\end{equation}
For the REUR we only need the diagonal entries of the off-diagonal blocks of the covariance matrix
\begin{equation}
    \begin{split}
        C^{\phi \pi}_{l l} &=\int \mathcal{D} \chi \, Q [\chi] \, \phi_{l} \pi_{l}\\ &= \sum_{\Vec{n}} \rho_{\Vec{n}}\left(\prod_{l'\neq l}\delta_{n'_{l'}n_{l'}}\right) \left( \frac{\Delta k}{2 \pi}\frac{\omega_l}{2 } \right)^{\frac{n_l+n'_l}{2}}  \int \mathrm{d} \pi_{l} \mathrm{d} \phi_{l} \frac{2 \pi }{\Delta k} \frac{i}{\omega_l}\left((i \pi_l)^{n'_l}\phi^{n_l}+\phi_l^{n'_l}(-i \pi_l)^{n_l}\right) \phi_{l} \pi_{l}  \\  
        &=  \frac{2 \pi }{\Delta k}\frac{i}{\omega_l}\,\rho^{(l)}_{00} \\ 
        &= \frac{2 \pi  }{\Delta k} \frac{i}{\omega_l}\left( 1 - \braket{n_l} \right),
    \end{split}
\end{equation}
where we have identified the total particle number per mode $l$ for fermions
\begin{equation}
    \braket{n_l} = \bTr \{ \boldsymbol{\rho} \,  \boldsymbol{a}^{\dagger}_l \boldsymbol{a}_l \} = \sum^1_{n_l=0} n_l \rho^{(l)}_{n_ln_l}.
\end{equation}


\bibliographystyle{quantum}
\bibliography{references.bib}

\end{document}